\documentclass[12pt]{article}
\usepackage{amssymb}
\usepackage{amsmath}
\usepackage{amsmath}
\usepackage{amsthm}
\usepackage{amsfonts}
\usepackage{graphicx}
\usepackage{enumerate}
\usepackage{bbm} 
\usepackage{dsfont}
\usepackage[authoryear,longnamesfirst]{natbib}
\usepackage{multirow}

\numberwithin{equation}{section}
\newtheorem{theorem}{Theorem}

\newtheorem{condition}{Condition}

\newtheorem{example}{Example}

\newtheorem{remark}{Remark}

\newtheorem{lemma}{Lemma}
\theoremstyle{definition}

\begin{document}

\title{\Large Testing Finite Moment Conditions for the Consistency and the \\ Root-N Asymptotic Normality of the GMM and M Estimators\thanks{First arXiv version: June 3, 2020. We thank Jia Li and Xiaoxia Shi for very useful comments. All the remaining errors are ours.}
}
\author{Yuya Sasaki\thanks{
Associate professor of economics, Vanderbilt University. 
Email: yuya.sasaki@vanderbilt.edu} \ and 
Yulong Wang\thanks{
Assistant professor of economics, Syracuse University. 
Email: ywang402@maxwell.syr.edu.}}
\date{}
%\date{First arXiv version: ...\\
%This version: ...}
\maketitle

\begin{abstract}\setlength{\baselineskip}{7.0mm}
Common approaches to inference for structural and reduced-form parameters in empirical economic analysis are based on the consistency and the root-n asymptotic normality of the GMM and M estimators.
The canonical consistency (respectively, root-n asymptotic normality) for these classes of estimators requires at least the first (respectively, second) moment of the score to be finite.
In this article, we present a method of testing these conditions for the consistency and the root-n asymptotic normality of the GMM and M estimators.
The proposed test controls size nearly uniformly over the set of data generating processes that are compatible with the null hypothesis.
Simulation studies support this theoretical result. 
Applying the proposed test to the market share data from the Dominick's Finer Foods retail chain, we find that a common \textit{ad hoc} procedure to deal with zero market shares in analysis of differentiated products markets results in a failure to satisfy the conditions for both the consistency and the root-n asymptotic normality.
\smallskip\\
{\bf JEL-Code:} C12
\\
\end{abstract}

%%%%%%%%%%%%%%%%%%%%%%%%%%%%%%%%%%%%%%%%%%%%%%%%%%%%%%%%%%%%%%%%%%%%%
\section{Introduction}
%%%%%%%%%%%%%%%%%%%%%%%%%%%%%%%%%%%%%%%%%%%%%%%%%%%%%%%%%%%%%%%%%%%%%

Many estimators of interest in economic analysis are GMM or M estimators.
They include, but are not limited to, the ordinary least squares (OLS) estimators, the generalized least squares (GLS) estimators, the quasi maximum likelihood estimators (QMLE), and the two stage least squares (2SLS) estimators.
Under random sampling, the consistency of these estimators is usually established via the weak law of large numbers (WLLN), which requires a finite first moment of the score.
The root-n asymptotic normality of these estimators is usually established via Lindeberg-L\'evy Central Limit Theorem (CLT), which requires a finite second moment of the score. %\footnote{Under non-identical distributions, versions of the relevant CLTs require even stronger conditions, such as Lindeberg's condition and Lyapunov's condition. \label{fn:slackness}}
These conditions are usually taken for granted by the authors of empirical economics papers that report point estimates and their standard errors, report confidence interval, and/or conduct hypothesis testing based on the limit normal distribution.

However, these assumptions are not necessarily plausibly satisfied in applications.
There are a couple of possible scenarios in which the score may not have finite first and second moments.
First, some dependent variables (e.g., infant birth weight\footnote{See \citet*[][Section 6.2]{ChernozhukovFernandezVal2011} for example.} and murder rate,\footnote{See \citet*[][Appendix A]{GandhiLuShi2017} for example.} as well as income, wealth, and stock returns) are reported to exhibit heavy tailed distributions, and their many outliers may contribute to heavy tailed distributions of residuals.
Second, suppose that a dependent variable is the logarithm of a variable, as is the case with demand analysis under differentiated products markets.
It is a common empirical practice to replace zeros by an infinitesimal value to avoid the logarithm of zero, but this operation may result in a heavy tailed distribution of the residuals for those observations with \textit{originally non-zero} values of the dependent variable -- we will later show a case in point based on actual empirical data.
%Both of these scenarios may entail infinite first and second moments of the score of an estimator under consideration.

In doubt about the assumptions of the finite first and second moments of the score in certain applications, we naturally desire to have a method of testing these conditions for the consistency and the root-n asymptotic normality.
This article is motivated by this objective, and we therefore propose a method of testing these assumptions.
With our proposed test, researchers can assess whether the bounded moment conditions for the consistency and the root-n asymptotic normality are satisfied for past and future empirical studies.
In the event where the test supports the consistency and the root-n asymptotic normality for a selected study, the test result will reinforce the credibility of the scientific conclusions reported by that study.
On the other hand, in the event where the test rejects the finite moment conditions for consistency or the root-n asymptotic normality for a selected study, researchers would like to substitute alternative robust methods -- see the related literature ahead.
In this way, our proposed method of testing the finite moment conditions is expected to contribute to enhancing the credibility of past and future empirical economic studies.

Our proposed method is based on extreme value theory, and works in the following simple manner.
Consider a test of finite $r$-th moment of the score -- set $r=1$ (respectively, $r=2$) for a test of the consistency (respectively, the root-n asymptotic normality).
First, sort the $r$-th power of the norm of the estimated score in descending order.
Second, pick the largest $k$ of these order statistics and self-normalize them.
We show that these self-normalized statistics asymptotically follow a known joint distribution up to the unknown tail index parameter.
A sub-unit (respectively, super-unit) value of this tail index parameter indicates a finite (respectively, infinite) $r$-th moment of the score.
Lastly, using these dichotomous characteristics, we construct a likelihood ratio test based on the limit joint distribution of the self-normalized statistics.
We establish a nearly uniform size control property of the proposed test over the set of data generating processes compatible with the null hypothesis of a finite $r$-th moment of the score.\footnote{See Section \ref{sec:method} ahead for a precise description of the near uniformity.}
This near uniformity property of the test is attractive since researchers do not \textit{ex ante} know or do not want to fix the true distribution of the $r$-th power of the norm of the score under the composite null hypothesis in consideration.

Simulation studies support the theoretical result of the nearly uniform size control property.
Applying the proposed method of testing to the widely used market share data from Dominick's Finer Foods retail chain, we find that the common \textit{ad hoc} treatment of zero market shares by adding an infinitesimal positive value results in a failure of the consistency and the root-n asymptotic normality.
This failure results from the fact that inclusion of logs of these infinitesimal numbers (i.e., large negative values) induces a heavy-tailed distribution of the regression residuals for observations with \textit{originally non-zero} market shares.

{\bf Relation to the Literature:}
We are not aware of any existing paper that develops a test of the finite moment condition for the consistency or the root-n asymptotic normality of the GMM or M estimators, as we do in this paper.
A different but related topic is a set of tools to test non- and weak-identification \citep*[e.g.,][]{Wright2003,StockYogo2005,InoueRossi2011,SandersonWindmeijer2016}.
These are related to our framework on one hand because non- and weak-identification also results in a failure of the canonical consistency and the root-n asymptotic normality, and therefore these testing methods serve for related objectives.
On the other hand, these are different from our framework because the non- and weak-identification concerns about non- and weak-invertibility of the expected gradient of the score,\footnote{For general matrix rank tests, see e.g., \citet{GillLewbel1992,CraggDonald1996,CraggDonald1997,RobinSmith2000,KleibergenPaap2006,CambaMendezKapetanios2009,AlSadoon2017}.} whereas the issue of our concern is instead about the finiteness of moments of the score as the conditions for the WLLN and CLT.
In this sense, the purposes of our method of test are different from those of the preceding methods of tests of non- and weak-identification, while they indeed play complementary roles.

Also related is the paper by \citet*{ShaoYuYu2001} that proposes a test of finite variance.
On the one hand, our test of the root-n asymptotic normality is also based on the test of finite second moments, similarly to \citet*{ShaoYuYu2001}.
On the other hand, our objective of testing the asymptotic normality for the GMM and M estimators requires to take into account that the score is not directly observed in data, but has to be estimated via the GMM or M estimation. 
With these similarities and differences, our proposed method also contributes to this existing literature on testing finite moments by allowing for generated data.

For scalar locations and single equation models, an infinite first or second moment of the score is often imputed to outliers.
Recognizing this issue, \citet*{Edgeworth1887} proposes to use the absolute loss instead of the square loss for a robust estimation of the equation parameters.
This idea later extends and generalizes to other robust methods based on the check losses \citep*{KoenkerBassett1978} and the Huber loss \citep*{Huber1992}.
While we propose a test of the finite second moment of a norm of the score for the root-n asymptotic normality of GMM and M estimators in general, there are existing papers that establish limit distribution theories (which are not necessarily root-n or normal) without requiring the finite second moment condition in these frameworks \citep*[e.g.,][]{DavisResnick1985,DavisResnick1986,DavisKnightLiu1992,HillProkhorov2016}.
In the event where our test fails to support the finite moment conditions, a researcher can resort to one of these alternative robust methods instead of relying on the standard methods of inference based on the consistency and the root-n asymptotic normality of the GMM and M estimators.

We in particular highlight the case of demand estimation in differentiated products markets with market share data as a motivating example, where the dependent variable in linear models is often defined as the logarithm of a variable that may occasionally take the value of zero.
Researchers sometimes substitute infinitesimal positive values for the zero in order to avoid the logarithm of zero, but this practice may also entail infinite first and second moments of the score -- see our empirical application in Section \ref{sec:application} ahead.
It is not the logarithm of these infinitesimal numbers \textit{per se} that act as outliers, but they induce a heavy-tailed distribution of residuals of observations with \textit{originally non-zero} market shares.
In light of these unfavorable test results for the common \textit{ad hoc} practice, we suggest that a researcher may in stead want to resort to alternative robust methods such as \citet*{GandhiLuShi2017} for demand analysis with zero market shares.
A similar suggestion applies to gravity analysis of international trade, where the logarithm of zero is ubiquitous -- a researcher may want to resort to alternative robust methods such as \citet*{SilvaTenreyro2006} for gravity analysis with zero trade flows.

Finally, our method is based on recent developments in extreme value theory. 
We refer readers to \citet*{HaanFerreira2006} for a very comprehensive review of this subject.
In particular, our inference approach is based on fixed-$k$ asymptotics, and takes advantage of and extends the technique developed by \citet*{MullerWang2017} and \citet*{Mueller2020}.
The fixed-$k$ approach is useful in practice, because the asymptotic size control is valid for any predetermined fixed number $k$, unlike traditional increasing-$k$ approaches that require a sequence of changing tuning parameters as the sample size grows for which a sensible choice rule is difficult to obtain in small samples.
The fixed-$k$ approach also allows for robustness against errors in preliminary estimation -- this type of robustness, benefiting from the fixed-tuning parameter setup, has been similarly explored in other contexts in the existing literature, e.g., the fixed-$b$ asymptotic inference under heteroskedasticity and autocorrelation proposed by \cite{KieferVogelsang2005} and the robust inference in kernel estimations proposed by \cite{CattaneoCrumpJansson2014}.
In constructing our likelihood ratio test, we take advantage of the computational algorithm developed by \citet*{ElliottMullerWatson2015}.

%%%%%%%%%%%%%%%%%%%%%%%%%%%%%%%%%%%%%%%%%%%%%%%%%%%%%%%%%%%%%%%%%%%%%
\section{Econometric Frameworks}\label{sec:frameworks}
%%%%%%%%%%%%%%%%%%%%%%%%%%%%%%%%%%%%%%%%%%%%%%%%%%%%%%%%%%%%%%%%%%%%%

In this section, we introduce the general frameworks of the GMM and M estimators for which we propose tests of the finite moment conditions for the consistency and the root-n asymptotic normality.
A concrete empirical example will follow after the presentation of the general frameworks.

%%%%%%%%%%%%%%%%%%%%%%%%%%%%%%%%%%%%%%%%%%%%%%%%%%%%%%%%%%%%%%%%%%%%%
\subsection{GMM and M Estimators}\label{sec:gmm_m}
%%%%%%%%%%%%%%%%%%%%%%%%%%%%%%%%%%%%%%%%%%%%%%%%%%%%%%%%%%%%%%%%%%%%%

{\bf M-Estimation:}
Consider the class of estimators defined by
\begin{align*}
\hat\theta = \arg\max_{\theta \in \Theta} \hat Q_n (\theta),
\end{align*}
where the criterion function $Q_n$ takes the form of
$
\hat Q_n(\theta) = n^{-1}\sum_{i=1}^n g_i(\theta).
$
Under regularity conditions for this class, the influence function representation takes the form of
\begin{align*}
\sqrt{n}\left(\hat\theta - \theta_0\right)
=
- \hat H_n(\theta_0)^{-1} \cdot \frac{1}{\sqrt{n}} \sum_{i=1}^n g_i'(\theta_0) + o_p(1),
\end{align*}
where
$
\hat H_n(\theta) = n^{-1}\sum_{i=1}^n D_\theta^2 g_i(\theta)
$
and
$
g_i'(\theta) = \nabla_\theta g_i(\theta).
$
The consistency of $\hat\theta$ (via the weak law of large numbers) requires 
$
E\left[\left\| g_i'(\theta_0) \right\|\right] < \infty.
$
Likewise, the asymptotic normality of $\sqrt{n}\left(\hat\theta - \theta_0\right)$ (via multivariate Lindeberg-L\'evy CLT) requires 
$
E\left[\left\| g_i'(\theta_0) \right\|^2\right] < \infty.
$
Common examples include the following two classes of estimators:
\begin{enumerate}
	\item (OLS) 
	$g_i'(\theta) = X_i \left(Y_i - X_i^\intercal \theta\right)$
	$\Rightarrow$
	$A_i^r(\theta) \equiv \left\| g_i'(\theta) \right\|^r = \{ \left(Y_i - X_i^\intercal \theta\right)^\intercal X_i^\intercal X_i \left(Y_i - X_i^\intercal \theta\right) \}^{r/2}$
	\item (QMLE) 
	$g_i'(\theta) = \nabla_\theta \ell(Y_i,X_i;\theta)$
	$\Rightarrow$
	$A_i^r(\theta) \equiv \left\| g_i'(\theta) \right\|^r = \{ \nabla_\theta \ell(Y_i,X_i;\theta)^\intercal \nabla_\theta \ell(Y_i,X_i;\theta) \}^{r/2}$
\end{enumerate}
In this paper, we propose tests of the null hypothesis: 
$
E\left[ A_i^1(\theta_0) \right] < \infty,
$
the condition that is required for establishing the consistency of $\hat\theta$; 
and the null hypothesis:
$
E\left[ A_i^2(\theta_0) \right] < \infty,
$
the condition that is required for establishing the asymptotic normality of $\sqrt{n}\left(\hat\theta - \theta_0\right)$.

\begin{remark}
Due to the related structures between M and Z estimators, our framework also applies to Z estimators.
\end{remark}

${}$\\
{\bf GMM:}
Next, consider the class of estimators defined by
\begin{align*}
\hat\theta = \arg\min_{\theta \in \Theta} \hat Q_n(\theta),
\end{align*}
where the criterion function $\hat{Q}_n$ takes the form of 
$
\hat Q_n(\theta) = \left[n^{-1} \sum_{i=1}^n g_i(\theta) \right]^\intercal \hat W \left[n^{-1} \sum_{i=1}^n g_i(\theta) \right].
$
Under regularity conditions for this class, the influence function representation takes the form of 
\begin{align*}
\sqrt{n}\left(\hat\theta - \theta_0\right) = -\left(\hat G_n(\hat\theta)^\intercal \hat W \hat G_n(\theta_0)\right)^{-1} \hat G_n(\hat\theta)^\intercal \hat W \frac{1}{\sqrt{n}} \sum_{i=1}^n g_i (\theta_0) + o_p(1),
\end{align*}
where
$
\hat G_n(\theta) = n^{-1} \sum_{i=1}^n \nabla_\theta g_i(\theta)
$
and
$
\hat W \stackrel{p}{\rightarrow} W_{0}.
$
The consistency of $\hat\theta$ (via the weak law of large numbers) requires 
$
E\left[\left\| g_i(\theta_0) \right\|\right] < \infty.
$
Likewise, the asymptotic normality of $\sqrt{n}\left(\hat\theta - \theta_0 \right)$ (via multivariate Lindeberg-L\'evy CLT) requires 
$
E\left[\left\| g_i(\theta_0) \right\|^2\right] < \infty.
$
A common example is:
\begin{enumerate}
	\item[3] (2SLS) 
	$g_i(\theta) = Z_i \left(Y_i - X_i^\intercal \theta\right)$
	$\Rightarrow$
	$A_i^r(\theta) \equiv \left\| g_i(\theta) \right\|^r = \{ \left(Y_i - X_i^\intercal \theta\right)^\intercal Z_i^\intercal Z_i \left(Y_i - X_i^\intercal \theta\right) \}^{r/2}$
\end{enumerate}
Similarly to the M-estimation case, we propose tests of the null hypothesis: 
$
E\left[ A_i^1(\theta_0) \right] < \infty,
$
which is required for establishing the consistency of $\hat\theta$; 
and the null hypothesis:
$
E\left[ A_i^2(\theta_0) \right] < \infty,
$
which is required for establishing the asymptotic normality of $\sqrt{n}\left(\hat\theta - \theta_0\right)$.

%%%%%%%%%%%%%%%%%%%%%%%%%%%%%%%%%%%%%%%%%%%%%%%%%%%%%%%%%%%%%%%%%%%%%
\subsection{Example: Demand Analysis in Differentiated Products Markets}
%%%%%%%%%%%%%%%%%%%%%%%%%%%%%%%%%%%%%%%%%%%%%%%%%%%%%%%%%%%%%%%%%%%%%

In applications, $A_i^r(\theta_0)$ may not have a finite moment in the presence of outliers in the dependent variable.
Outliers may be innate in data in some applications.
In other applications, outliers may be produced as artifacts of \textit{ad hoc} procedures taken by researchers.
%For instance, when a dependent variable is the logarithm of a variable, researchers often substitute positive values $\Delta$ for zeros in order to avoid the logarithm of zero.
%When $\Delta$ is too small, however, it may induce a heavy tailed distribution in regression residuals.
In the empirical application to be presented in Section \ref{sec:application} ahead, we highlight this point in the context of the demand analysis under the following setup.

\begin{example}[Demand Analysis]\label{example:blp}
  Demand estimation with market share data\footnote{This framework is drawn from the literature on the estimation of demand for differentiated products \citep*{Berry1994,BerryLevinsohnPakes1995}. See \citet*{AckerbergBenkardBerryPakes2007} for a survey.} in the logit case is based on GMM with the moment function defined by 
	$$
	g_{jt}(\theta) = Z_{jt} \left(\ln(S_{jt}) - \ln(S_{0t}) - P_{jt} \theta_{1} - X_{it}^\intercal \theta_{-1}\right), 
	$$
	where $j$ indexes products, $t$ indexes markets, $S_{jt}$ denotes the share of product $j$ in market $t$, $P_{jt}$ denotes the price, $X_{jt}$ denotes product characteristics, and $Z_{jt}$ denotes instruments.
	In this setting, 
	$$
	A_{jt}^r(\theta) = \{ \left(\ln(S_{jt}) - \ln(S_{0t}) - P_{jt} \theta_{1} - X_{it}^\intercal \theta_{-1}\right)^\intercal Z_{jt}^\intercal Z_{jt} \left(\ln(S_{jt}) - \ln(S_{0t}) - P_{jt} \theta_{1} - X_{it}^\intercal \theta_{-1}\right) \}^{r/2},
	$$
	evaluated at $\theta=\theta_0$, may not have a finite moment when $\ln(S_{jt})$ has a heavy tailed distribution.
	Furthermore, in the presence of zero market shares, there is a common \text{ad hoc} practice of replacing zero shares by infinitesimal shares in order to avoid the log of zeros, but only to artificially produce heavy tailed distribution of regression residuals for observations with ``{originally non-zero}'' market shares.
	Such a practice can result in a failure of the consistency and the root-n asymptotic normality -- see our empirical applications in Section \ref{sec:application}.
	In the event where one rejects the finite moment conditions for consistency or the root-n asymptotic normality, a researcher can still resort to robust inference procedures \citep*[e.g.,][]{GandhiLuShi2017} specialized in this empirical framework.
\end{example}

%%%%%%%%%%%%%%%%%%%%%%%%%%%%%%%%%%%%%%%%%%%%%%%%%%%%%%%%%%%%%%%%%%%%%
\section{The Test}\label{sec:method}
%%%%%%%%%%%%%%%%%%%%%%%%%%%%%%%%%%%%%%%%%%%%%%%%%%%%%%%%%%%%%%%%%%%%%

This section presents an overview of the procedure of our proposed test.
Formal theoretical justifications for why this method works will be presented in Section \ref{sec:theory}.

First, consider a non-negative random variable $A_i^r(\theta_{0})$, examples of which are introduced in Section \ref{sec:frameworks}. 
Whether its moment is finite is fully determined by its right-tail behavior. 
Specifically, suppose that there is a single parameter $\xi \ge 0$ so that $n^{\xi}$ characterizes the order of magnitude of the sample maximum of $A_i^r(\theta_{0})$. Then, by extreme value theory, the finiteness of the moment of $A_i^r(\theta_{0})$ is equivalent to the condition that $\xi < 1$. 
To obtain a compact null space, we set $\varepsilon$ to be some small slackness parameter, say 0.01.\footnote{Another possible treatment is to switch the null and alternative hypotheses so that $H_{0}$ is $\xi \in [1, \bar{\xi}$]. However, deriving the asymptotic behavior of our test will be challenging under such $H_{0}$, if possible at all, since the estimator of $\theta_{0}$ could behave poorly.}
Then, we write the two competing hypotheses:
\begin{equation}
H_{0}:\xi \in \lbrack 0,1-\varepsilon ] \text{ against }
H_{1}:\xi \in \left( 1-\varepsilon ,\bar{\xi} \right] \text{,}  \label{hypo}
\end{equation}
where $\bar{\xi}$ is the upper bound of the parameter space that includes all empirically relevant values of $\xi$. We set $\bar{\xi}=2$ in later sections.

Since the limiting density (see (\ref{ev_pdf}) below) and the power function are continuous in $\xi$, setting the slackness parameter $\varepsilon$ is innocuous and without loss of generality unlike the seemingly related treatments in the contexts of the (near) unit root or weak-/non-identification. 
In practice, we can set $\varepsilon$ arbitrarily close to zero (and $\bar{\xi}$ arbitrarily large).
In this sense, the near uniformity of our proposed test means controlling size uniformly over $\xi \in \lbrack 0,1-\varepsilon ]$, which can be arbitrarily close to $\lbrack 0,1)$. 
%Though our near uniformity does not cover all distributions of $A_i^r(\theta_{0})$ that posses a finite expectation, it is suitable in the following two senarios. 
In other words, it guarantees the size control over all distributions of $A_i^r(\theta_{0})$ that possess a finite $1+\delta$ moment for $\delta < \varepsilon/(1-\varepsilon)$. 
%This is relevant for the CLTs under non-identical distributions. 
As recently studied by \cite{Mueller2020}, the asymptotic normality may provide a very poor approximation for both the classic t-statistic and the bootstrap if the second moment is finite but the third moment is not. 
%Therefore, if one sets $r=1$, then rejecting $H_0$ implies that the $1+\delta$ moment condition is violated, which jeopardizes the consistency of $\hat\theta$ for $\theta_0$.
If one sets $r=2$, then rejecting $H_0$ indicates that the asymptotic normality of $\sqrt{n}\left(\hat\theta-\theta_0\right)$ is not reliable.

Second, in order to test the above hypothesis, we would like to observe large values of $A_i^r(\theta_{0})$, but it is of course unobserved. 
To overcome this issue, we make use of a consistent estimator $\hat\theta$ of $\theta_0$.
Let 
$
A_{(1)}^r(\hat{\theta})\geq A_{(2)}^r(\hat{\theta})\geq \ldots \geq A_{(n)}^r(\hat{\theta})
$ 
denote the order statistics by sorting $\{A_i^r(\hat\theta)\}_{i=1}^n$ in the descending order.
For a pre-determined integer $k \ge 3$, collect the $k$ order statistics as
\begin{equation*}
\mathbf{A}^r\left( \hat{\theta}\right) =\left[ A_{(1)}^r(\hat{\theta}),A_{(2)}^r(\hat{\theta}),\ldots ,A_{(k)}^r(\hat{\theta})\right] ^{\intercal }.
\end{equation*}
By extreme value theory again, the joint distribution of the largest order statistics asymptotically approaches a well-defined parametric joint distribution that is fully characterized by the location, the scale, and the scalar parameter $\xi$. 
Therefore, if we conduct location and scale normalization by considering the statistics
\begin{equation*}
\mathbf{A}_{\ast}^r\left( \hat{\theta}\right) = \frac{\mathbf{A}^r\left(\hat{\theta}\right) -A_{\left( k\right) }^r \left( \hat{\theta}\right) }{A_{\left( 1\right) }^r \left( \hat{\theta}\right) -A_{\left( k\right) }^r \left(\hat{\theta}\right) },
\end{equation*}
then $\mathbf{A}_{\ast}^r(\hat{\theta})$ converges in distribution to a limiting random vector $\mathbf{V}_{\ast}$, whose density $f_{\mathbf{V}_{\ast}}$ is fully characterized by $\xi$ and is invariant to location, scale, and the order $r$ -- see (\ref{ev_pdf}) ahead for the formula of $f_{\mathbf{V}_{\ast}}$. 
The estimation error in $\hat\theta$ is asymptotically negligible since it is of a smaller order of magnitude than the largest order statistics of $A_i^r(\theta_{0})$ under the null hypothesis.

Now, the limiting testing problem has become straightforward: we construct a test based on a random draw of $\mathbf{V}_{\ast}$ from its parametric density $f_{\mathbf{V}_{\ast}}$ about the only unknown scalar parameter $\xi$. 
When the null and alternative hypotheses are both simple, the optimal solution is known to be the Neyman-Pearson test, where large values of the likelihood ratio statistic reject the null hypothesis. 
Therefore, we transform the null and alternative hypotheses of (\ref{hypo}) into simple ones by considering weighted average likelihoods, and our proposed test rejects the null hypothesis that $A_i^r(\theta_0)$ has a finite moment if
\begin{equation*}
\frac{\int f_{\mathbf{V}_{\ast }}\left(\mathbf{A}_{\ast}^r(\hat{\theta}) ;\xi\right) dW\left( \xi \right) }{\int f_{\mathbf{V}_{\ast}}\left(\mathbf{A}_{\ast}^r(\hat{\theta}) ;\xi \right) d\Lambda \left( \xi \right) }>1,  
\end{equation*}
where $W\left( \cdot \right)$ denotes a weight chosen to reflect the importance of rejecting different alternatives, and 
$\Lambda \left( \cdot \right) $ is some pre-determined weight defined on the null space. 
The critical value is subsumed in $\Lambda \left( \cdot \right) $ so that the likelihood ratio on the left-hand side is compared with one on the right-hand side. 
More details about this test are presented in the following section.

\begin{figure}
	\centering
		\includegraphics[width=1.00\textwidth]{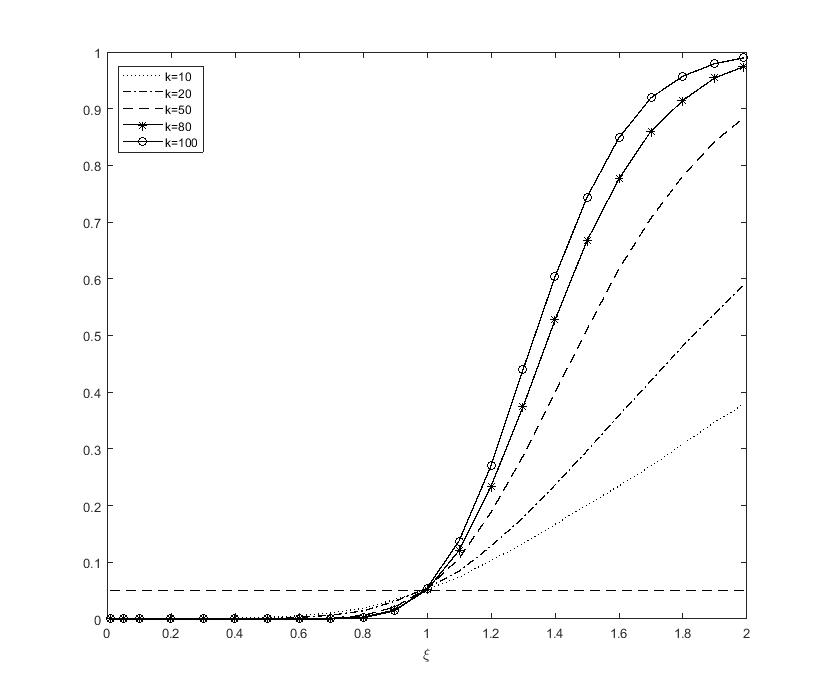}
	\caption{Rejection probabilities of the test with $\mathbf{V}_{\ast }$ generated from  $f_{\mathbf{V}_{\ast }}$ with $\xi\in [0,2]$ and the nominal size of $0.05$. The plots are based on numerical simulations with 10000 iterations.}
	\label{fig:power_curves}
	${}$
\end{figure}

We close the preview with some heuristic discussions of the \textit{asymptotic} property of the new test - formal discussions will follow in Section \ref{sec:theory}. Figure \ref{fig:power_curves} plots oracle rejection probabilities of the test with $\mathbf{V}_{\ast }$ generated from the limiting distribution $f_{\mathbf{V}_{\ast }}$ with various values of $\xi$ and the nominal size of $0.05$. 
The plots are based on simulations with 10000 iterations.
(This is a power prescription and is different from Monte Carlo simulation studies. Full-blown Monte Carlo simulation studies with concrete econometric models and small samples will be conducted and presented in Section \ref{sec:simulation} to evaluate the \textit{finite sample} performance.)
Observe that the rejection probabilities for $\xi \in [0,1-\varepsilon]$ are uniformly dominated by the nominal size, $0.05$.
In other words, the test has a size control property for nearly all distributions with tail index less than one, where the slackness $\varepsilon$ can be made arbitrarily small. 
This uniformity property is useful in practice because researchers do not \textit{ex ante} know or do not want to fix the true value of $\xi$ in econometric models of their interest in the composite null hypothesis of consideration.

%Second, observe that the power of the test increases in $k$ for any fixed $\xi > 1$. We can formalize the consistency of the test by setting $k \rightarrow \infty $ and $k/n \rightarrow 0$. Then the largest $k$ order statistics become asymptotically independent, and we can construct consistent estimators of $\xi$ by conducting the maximum likelihood estimation \citep[e.g.,][]{Hill1975,Smith1987}. Based on the asymptotic normality of these estimators, we can consistently test against any fixed alternative. In this sense, our fixed-$k$ asymptotic framework is suitable for a local asymptotic analysis where our likelihood ratio test achieves the optimal local power (with respect to the weighted alternative) as implied by the Neyman-Pearson lemma.

The following section formally presents the asymptotic theory for the test.

%%%%%%%%%%%%%%%%%%%%%%%%%%%%%%%%%%%%%%%%%%%%%%%%%%%%%%%%%%%%%%%%%%%%%
\section{Asymptotic Theory}\label{sec:theory}
%%%%%%%%%%%%%%%%%%%%%%%%%%%%%%%%%%%%%%%%%%%%%%%%%%%%%%%%%%%%%%%%%%%%%

We now present a formal theory to guarantee that our proposed test works in large samples. 
For the convenience of exposition, we first introduce additional notations and some definitions.

Denote by $D_{i}$ the $i$-th observation so that we can write $A_{i}^r\left( \theta \right) = A^r\left( \theta ;D_{i}\right)$. 
For example, $D_i = (X_i^\intercal,Y_i)^\intercal$ in the context of the OLS, and $D_i = (X_i^\intercal,Z_i^\intercal,Y_i)^\intercal$ in the context of the 2SLS presented in Section \ref{sec:gmm_m}.
Let $F_{A^r\left( \theta \right) }$ denote the cumulative distribution function (CDF) of $A_{i}^r\left( \theta \right)$ and $\theta _{0}$ denote the (pseudo-) true value of $\theta $. 
Let $B_{\eta _{n}}\left( \theta _{0}\right) $ denote an open ball centered at $\theta _{0}$ with radius $\eta _{n}\rightarrow 0$. 
Let $f_{A^r\left( \theta _{0}\right) }$ and $Q_{A^r\left( \theta _{0}\right) }$ be the probability density function (PDF) and quantile function of $A_{i}^r\left( \theta _{0}\right) $, respectively.

We say that a distribution $F$ is within the domain of attraction of the extreme value distribution, denoted by $F \in \mathcal{D}\left( G_{\xi}\right)$, if there exist sequences of constants $a_n$ and $b_n$ such that for every $v$,
$$
\lim_{n \rightarrow \infty} F^{n}(a_n v + b_n) = G_\xi(v)
$$
holds,
where
$$
G_\xi(v) =
\begin{cases}
\exp\left( - (1 + \xi v)^{-1/\xi} \right) & 1 + \xi v > 0, \xi \neq 0
\\
\exp\left(-e^{-v}\right) & v \in \mathbb{R}, \xi=0
\end{cases}
$$
is referred to as the generalized extreme value distribution. This condition, characterizing the tail shape of the underlying distribution, is mild and satisfied by many commonly used distributions. 
In particular, the case with $\xi>0$ covers distributions with a regularly varying tail, including, for example, Pareto, Student-t, and F distributions. 
The case with $\xi=0$ covers thin tailed distributions, such as the Gaussian family. 
The case with $\xi<0$ covers distributions with bounded right end-point, such as the uniform distribution. 
See \citet*[][Ch.1]{HaanFerreira2006} for a comprehensive review.
With these notations and definitions, we impose the following regularity conditions to prove the uniform size control property of our proposed test. 

%%%%%%%%%%%%%%%%%%%%%%%%%%%%%%%%%%%%%%%%%%%%%%%%%%%%%%%%%%%%%%%%%%%%%
\begin{condition}\label{conditions}
The following conditions are satisfied.
\begin{enumerate}[(i)]
\item $D_{i}$ is i.i.d. from some underlying distribution that does not change with $n$.

\item $F_{A^r\left( \theta _{0}\right) }\in \mathcal{D}\left( G_{\xi} \right) $ with $\xi \geq 0$ and $Q_{A^r\left( \theta_{0}\right) }\left( 1\right) =\infty $.

\item $\hat{\theta}-\theta _{0}=o_{p}\left( 1\right) $. 
For some $\eta_{n} \rightarrow 0$, if $\xi>0$, 
then $\sup_{i}\sup_{\theta \in B_{\eta _{n}}\left( \theta _{0}\right)}\left\vert \left\vert \frac{\partial A_{i}^r\left( \theta \right) }{\partial\theta }\right\vert \right\vert =O_{p}\left( n^{\xi}\right) $;
and if $\xi=0$, 
then $\sup_{i}\sup_{\theta \in B_{\eta _{n}}\left(\theta _{0}\right) }\left\vert \left\vert \frac{\partial A_{i}^r\left( \theta\right) }{\partial \theta }\right\vert \right\vert =O_{p}\left( nf_{A^r\left(\theta _{0}\right) }\left( Q_{A^r\left( \theta _{0}\right) }\left(1-1/n\right) \right) \right)$.
\end{enumerate}
\end{condition}
%%%%%%%%%%%%%%%%%%%%%%%%%%%%%%%%%%%%%%%%%%%%%%%%%%%%%%%%%%%%%%%%%%%%%

Condition \ref{conditions} (i) requires random sampling from some fixed population distribution. 
It rules out the trivial case where the location and the scale of the data diverge with the sample size. 
As a consequence, the existence of the moment only depends on the tail heaviness of the underlying distribution, which should remain invariant to location and scale shifts.

Condition \ref{conditions} (ii) requires that the distribution of $A^r(\theta_0)$ falls in the domain of attraction of the extreme value distribution, and that it has an unbounded support.\footnote{The finiteness of $\mathbb{E}[A_i^r(\theta_0)]$ is trivially satisfied if $\xi$ is negative.}
This domain of attraction assumption bridges the finiteness of moments and the tail heaviness. 
In particular, the finiteness of the first moment of $A_i^r(\theta_0)$ is equivalent to the condition that $\xi$ is less than than $1$ \citep*[cf.][Ch.5.3.1]{HaanFerreira2006}.\footnote{Besides, if $A_i^{1}(\theta_0)$ satisfies Condition \ref{conditions} (ii) with the tail index $\xi>0$, then $A_i^{2}(\theta_0)$ satisfies it with the tail index $2\xi$. See, for example, \citet*[][Proposition B.1.9]{HaanFerreira2006}.}  
Therefore, our hypothesis testing problem can be written as in (\ref{hypo}).

The first part of Condition \ref{conditions} (iii) requires that the estimator $\hat\theta$ is consistent for $\theta_0$.
Note that this in general only requires the identification and finite first moments of the score.
For the case of testing the finite first moment condition for consistency (i.e., the case of setting $r=1$), this assumption is satisfied under the null hypothesis in (\ref{hypo}).
For the case of testing the finite second moment condition for root-n asymptotic normality (i.e., the case of setting $r=2$), this assumption is satisfied even under a range of alternative hypotheses as well as under the null hypothesis.
In any case, the fact that this consistency condition is satisfied under the null hypothesis allows us to establish a size control property based on this condition.

The second part of Condition \ref{conditions} (iii) requires that the gradient of $A_i^r(\theta)$ grows not too fast as the sample size increases.
Since this last piece of the condition is a high-level statement, it will be useful to consider stronger lower level sufficient conditions in a specific example.
Since we focus on the case of the GMM in our application, we look at a lower level condition in the context of the GMM.

%%%%%%%%%%%%%%%%%%%%%%%%%%%%%%%%%%%%%%%%%%%%%%%%%%%%%%%%%%%%%%%%%%%%%
${}$\\
{\bf Discussion of Condition \ref{conditions} (iii) -- Case of GMM:}
Consider the case of setting $r=2$ for testing the root-n asymptotic normality.
Recall that we define
$
A_{i}^2\left( \theta \right) =\left( Y_{i}-X_{i}^{\intercal }\theta\right) ^{\intercal }Z_{i}^{\intercal }Z_{i}\left( Y_{i}-X_{i}^{\intercal}\theta \right).$
Thus,
\begin{eqnarray*}
\frac{\partial A_{i}^2\left( \theta \right) }{\partial \theta }
&=&-2X_{i}Z_{i}^{\intercal }Z_{i}\left( Y_{i}-X_{i}^{\intercal }\theta\right) \\
&=&-2X_{i}Z_{i}^{\intercal }Z_{i}u_{i}+2X_{i}Z_{i}^{\intercal}Z_{i}X_{i}^{\intercal }\left( \hat{\theta}-\theta \right).
\end{eqnarray*}
The triangle inequality and Cauchy-Schwartz inequality yield
\begin{eqnarray}
\sup_{i}\sup_{\theta \in B_{\eta_{n} }\left( \theta _{0}\right) }\left\vert\left\vert \frac{\partial A_{i}^2\left( \theta \right) }{\partial \theta }\right\vert \right\vert \leq 2\sup_{i}\left\vert \left\vert X_{i}Z_{i}^{\intercal }Z_{i}u_{i}\right\vert \right\vert+2\sup_{i}\left\vert \left\vert X_{i}Z_{i}^{\intercal }Z_{i}X_{i}^{\intercal}\right\vert \right\vert \cdot \sup_{\theta \in B_{\eta _{n}}\left( \theta
_{0}\right) }\left\vert \left\vert \hat{\theta}-\theta \right\vert \right\vert \ \notag \\
\leq 2\sup_{i}\left\vert \left\vert X_{i}Z_{i}^{\intercal }\right\vert\right\vert \left( \sup_{i}\left\vert \left\vert A_{i}^2\left( \theta_{0}\right) \right\vert \right\vert \right) ^{1/2}+2\left(\sup_{i}\left\vert \left\vert X_{i}Z_{i}^{\intercal }\right\vert \right\vert\right) ^{2}\cdot \sup_{\theta \in B_{\eta _{n}}\left( \theta _{0}\right)}\left\vert \left\vert \hat{\theta}-\theta \right\vert\right\vert.  \label{dA decom}
\end{eqnarray}
Condition \ref{conditions} (ii) implies that 
$
\sup_{i}\left\vert \left\vert A_{i}^2\left(\theta _{0}\right) \right\vert \right\vert \rightarrow \infty 
$ 
so that 
$
\left( \sup_{i}\left\vert \left\vert A_{i}^2\left( \theta _{0}\right) \right\vert \right\vert \right) ^{1/2}
$ 
is of a smaller order than 
$
\sup_{i}\left\vert \left\vert A_{i}^2\left( \theta _{0}\right) \right\vert \right\vert. 
$ 
Therefore, a sufficient condition for Condition \ref{conditions} (iii) is that 
$
\sup_{i}\left\vert \left\vert X_{i}Z_{i}^{\intercal }\right\vert \right\vert 
$ 
is of a smaller order than 
$
\left( \sup_{i}\left\vert \left\vert A_{i}^2\left( \theta _{0}\right)
\right\vert \right\vert \right) ^{1/2}.
$ 
An even stronger sufficient condition for this sufficient condition is that $X_{i}$ and $Z_{i}$ have bounded supports, which is satisfied by the typical applications in the demand analysis, in particular the one that we consider in our empirical application in Section \ref{sec:application}. 
$\square$
\\${}$
%%%%%%%%%%%%%%%%%%%%%%%%%%%%%%%%%%%%%%%%%%%%%%%%%%%%%%%%%%%%%%%%%%%%%

As discussed earlier, the domain of attraction assumption as in Condition \ref{conditions} (ii) allows for the hypothesis testing problem to be written as in (\ref{hypo}). 

Recall the following notations from Section \ref{sec:method}. Let 
$
A_{(1)}^r(\hat{\theta})\geq A_{(2)}^r(\hat{\theta})\geq \ldots \geq A_{(n)}^r(\hat{\theta})
$ 
denote the order statistics by sorting $\{A_i^r(\hat\theta)\}_{i=1}^n$ in the descending order, and
\begin{equation*}
\mathbf{A}^r\left( \hat{\theta}\right) =\left[ A_{(1)}^r(\hat{\theta}),A_{(2)}^r(\hat{\theta}),\ldots ,A_{(k)}^r(\hat{\theta})\right] ^{\intercal }.
\end{equation*}
The following lemma shows that these order statistics asymptotically follow the joint extreme value distribution.

%%%%%%%%%%%%%%%%%%%%%%%%%%%%%%%%%%%%%%%%%%%%%%%%%%%%%%%%%%%%%%%%%%%%%
\begin{lemma}\label{lemma:ev}
Under Condition \ref{conditions}, there exist sequences of constants $a_n$ and $b_n$ depending on $F_{A^r \left( \theta_{0} \right) }$ such that, for any fixed $k$,
\begin{equation*}
\frac{\mathbf{A}^r\left( \hat{\theta}\right) -b_{n}}{a_{n}}\overset{d}{%
\rightarrow }\mathbf{V}\equiv \left( V_{1},...,V_{k}\right) ^{\intercal },
\end{equation*}%
where $\mathbf{V}$ is jointly distributed with the density given by $f_{\mathbf{V}|\xi }(v_{1},...,v_{k})=G_{\xi }(v_{k})\prod_{i=1}^{k}g_{\xi
}(v_{i})/G_{\xi }(v_{i})$ on $v_{k}\leq v_{k-1}\leq \ldots \leq v_{1}$, and $%
g_{\xi }(v)=\partial G_{\xi }(v)/\partial v$.
\end{lemma}
%%%%%%%%%%%%%%%%%%%%%%%%%%%%%%%%%%%%%%%%%%%%%%%%%%%%%%%%%%%%%%%%%%%%%

A proof is provided in Appendix \ref{sec:lemma:ev}.

Since $a_{n}$ and $b_{n}$ are unknown, we would like to eliminate them in constructing feasible test statistics.
We do so by constructing the self-normalized statistic
\begin{equation*}
\mathbf{A}_{\ast}^r\left( \hat{\theta}\right) = \frac{\mathbf{A}^r\left(\hat{\theta}\right) -A_{\left( k\right) }^r\left( \hat{\theta}\right) }{A_{\left( 1\right) }^r\left( \hat{\theta}\right) -A_{\left( k\right) }^r\left(\hat{\theta}\right) }.
\end{equation*}%
By the continuous mapping theorem, change of variables, and Lemma \ref{lemma:ev}, we obtain
\begin{equation*}
\mathbf{A}_{\ast}^r\left( \hat{\theta}\right) \overset{d}{\rightarrow }%
\mathbf{V}_{\ast}\equiv \frac{\mathbf{V}-V_{k}}{V_{1}-V_{k}},
\end{equation*}%
where the density function $f_{\mathbf{V}_{\ast }}$ of the limit observation $\mathbf{V}_\ast$ is given by
\begin{equation}
f_{\mathbf{V}_{\ast }}\left( \mathbf{v}_{\ast };\xi \right) =\Gamma \left(
k\right) \int_{0}^{\infty }s^{k-2}\exp \left( \left( -1-1/\xi \right)
\sum_{i=1}^{k}\log \left( 1+\xi v_{\ast i}s\right) \right) ds,
\label{ev_pdf}
\end{equation}%
and $v_{\ast i}$ denotes the $i$-th component of $\mathbf{v}_{\ast }$.
With this density function, we construct the likelihood ratio test
\begin{equation}
\varphi \left( \mathbf{V}_{\ast }\right) 
=
\mathbf{1}
\left[ \frac{\int_{1-\varepsilon }^{2}f_{\mathbf{V}_{\ast }}\left( \mathbf{V}_{\ast };\xi\right) dW\left( \xi \right) }{\int_{0}^{1-\varepsilon }f_{\mathbf{V}_{\ast
}}\left( \mathbf{V}_{\ast };\xi \right) d\Lambda \left( \xi \right) }>1\right] ,  \label{test}
\end{equation}
where $W\left( \cdot \right)$ denotes a weight chosen to reflect the importance of rejecting different alternatives\footnote{We set $W$ to be the uniform distribution on $[0,1-\varepsilon]$ with $\varepsilon=0.01$ in later sections.} and $\Lambda \left( \cdot \right) $ is some pre-determined weight that subsumes the critical value and transforms the composite null space into a simple one. Besides, such $\Lambda$ is referred to as the least favorable distribution \citep*[e.g.,][Ch.3.8]{LehmannRomano2005} that guarantees the uniform size control. \citet*{ElliottMullerWatson2015} develop a generic algorithm to numerically construct $\Lambda$, which we adapt to our setting -- see Appendix \ref{sec:computational_algorithm} for details. 

The following theorem establishes the asymptotic uniform size control of our test (\ref{test}), which is the main theoretical result of this article. 
Let $\mathbb{E}_{\xi }\left[\; \cdot \; \right]$ denote the expectation with respect to the density (\ref{ev_pdf}) with the parameter value $\xi$.

\begin{theorem}\label{theorem:main}
Suppose that Condition \ref{conditions} holds. 
For any fixed $k$, 
\begin{equation*}
\lim_{n\rightarrow \infty }\sup_{\xi \in [ 0,1-\varepsilon ] }\mathbb{E}_{\xi }\left[ \varphi \left( \mathbf{A}_{\ast }^r ( \hat{\theta} ) \right) \right] \leq \alpha.
\end{equation*}
\end{theorem}

A proof is provided in Appendix \ref{sec:theorem:main}. 

A few remarks are in order about this new test.
First, the fixed-$k$ asymptotic design leads to the desired uniform size control property as stated in the theorem. 
This feature provides a practical advantage because a researcher does not \textit{ex ante} know or does not want to fix the true distribution under the composite null hypothesis.
Second, the fixed-$k$ design is useful not only for the uniform size control but also for the robustness of the test against estimation error in $\hat\theta$.
This is because the fixed-$k$ design allows for the estimation error, $\hat{\theta}-\theta_0$, to be $o_p(1)$ and hence to be dominated by the infeasible largest order statistics of $\{ A_i^r({\theta_0})\}$. 
This type of robustness, benefiting from the fixed-tuning parameter setup, has been similarly explored in other contexts in the existing literature. 
See, for example, the fixed-$b$ asymptotic inference under heteroskedasticity and autocorrelation proposed by \cite{KieferVogelsang2005} and the robust inference in kernel estimations proposed by \cite{CattaneoCrumpJansson2014}.

While we focus on the fixed-$k$ asymptotic design for these practical and theoretical advantages, it is also possible to analyze an increasing-$k$ asymptotic design.
Specifically, if we consider the asymptotic setting where $k \rightarrow \infty$ and $k/n \rightarrow 0$, then $\xi$ can be consistently estimated -- many such estimators exist in the statistics literature \citep[see, for example,][Chapter 3]{HaanFerreira2006}. 
These estimators are usually asymptotically normal with the root-$k$ convergence rate.
One can thereby build a confidence interval for $\xi$ to conduct a test of $H_0$ for $r=2$ that is consistent against fixed alternatives.
Furthermore, we can always numerically examine the asymptotic power of the test as in Figure \ref{fig:power_curves} presented in Section \ref{sec:method}, which shows that the test has good power as long as $k$ is not too small.
With this device, a practitioner can implement preliminary power analysis in the choice of $k$ before actually conducting the test, and this feature is practically more useful in finite samples.

%%%%%%%%%%%%%%%%%%%%%%%%%%%%%%%%%%%%%%%%%%%%%%%%%%%%%%%%%%%%%%%%%%%%%
\section{Simulation Studies}\label{sec:simulation}
%%%%%%%%%%%%%%%%%%%%%%%%%%%%%%%%%%%%%%%%%%%%%%%%%%%%%%%%%%%%%%%%%%%%%

Using Monte Carlo simulations, we demonstrate that the proposed test has the claimed uniform size control property.
We consider two of the most popular econometric models, namely the linear regression model and the linear IV model, for data generating designs.
For each of these two designs, we consider the test of the consistency (by setting $r=1$) and the test of the root-n asymptotic normality (by setting $r=2$).

%%%%%%%%%%%%%%%%%%%%%%%%%%%%%%%%%%%%%%%%%%%%%%%%%%%%%%%%%%%%%%%%%%%%%
\subsection{Simulation Setup}
%%%%%%%%%%%%%%%%%%%%%%%%%%%%%%%%%%%%%%%%%%%%%%%%%%%%%%%%%%%%%%%%%%%%%

First, consider the linear regression model:
\begin{align*}
Y_i = & \theta_1 + \theta_2 X_i + U_i,
\end{align*}
where $\theta = (\theta_1,\theta_2)^\intercal = (1,1)^\intercal$.
The independent variable is generated according to $X_i \sim N(0,1)$.
The error $U_i$ is generated according to the zero symmetric Pareto distribution with tail index $\xi_U$, independently from $X_i$.
We vary the value of $\xi_U$ across sets of simulations.
The test is based on the $r$-th moment of the score:
$
A_i^r(\theta) = (1 + X_i^2)^{r/2} \cdot |Y_i - \theta_1 - \theta_2 X_i |^r
$
for $r=1,2$.
Since we do not know $\theta_0$, we replace $\theta_0$ by the OLS $\hat\theta$.
We thus use $k$ order statistics of
\begin{align*}
A_i^r(\hat\theta) = (1 + X_i^2)^{r/2} \cdot |Y_i - \hat\theta_1 - \hat\theta_2 X_i |^r
\end{align*}
to construct our test, following the procedure outlined in Section \ref{sec:method}.
We set $r=1$ for the test of the consistency of $\hat\theta$, and $r=2$ for the test of the asymptotic normality of $\sqrt{n}\left(\hat\theta-\theta_0\right)$.

Second, consider the linear IV model:
\begin{align*}
Y_i =& \theta_1 + \theta_2 X_i + U_i + V_i
\\
X_i =& \pi_1 + \pi_2 Z_i + R_i
\end{align*}
where $\theta = (\theta_1,\theta_2)^\intercal = (1,1)^\intercal$ and $\pi = (\pi_1,\pi_2)^\intercal = (1,1)^\intercal$.
The instrument is generated according to $Z_i \sim N(0,1)$ independently of the tri-variate error components $(U_i,V_i,R_i)^\intercal$.
The heavy tailed part of the error $U_i$ is generated according to the zero symmetric Pareto distribution with tail index $\xi_U$, independently from $(V_i,R_i)^\intercal$.
We vary the value of $\xi_U$ across sets of simulations.
The endogenous part of the error components $(V_i,R_i)^\intercal$ is generated according to 
$(V_i,R_i)^\intercal \sim N(\vec{0},\Sigma)$ where $\Sigma = (1,0.5;0.5,1)$.
The test is based on the $r$-th moment of the score:
$
A_i^r(\theta) = (1 + Z_i^2)^{r/2} \cdot |Y_i - \theta_1 - \theta_2 X_i |^r
$
for $r=1,2$.
Since we do not know $\theta_0$, we replace $\theta_0$ by the IV estimator $\hat\theta$.
We thus use $k$ order statistics of
\begin{align*}
A_i^r(\hat\theta) = (1 + Z_i^2)^{r/2} \cdot |Y_i - \hat\theta_1 - \hat\theta_2 X_i |^r
\end{align*}
to construct our test, following the procedure outlined in Section \ref{sec:method}.
We set $r=1$ for the test of the consistency of $\hat\theta$, and $r=2$ for the test of the asymptotic normality of $\sqrt{n}\left(\hat\theta-\theta_0\right)$.

For each of the linear regression model and the linear IV model introduced above, we experiment with sample sizes of $n= 10^4$, $10^5$ and $10^6$, which are similar to the sample size that we actually encounter in our empirical application in Section \ref{sec:application}.
For testing the finite first moment condition for the consistency of $\hat\theta$, we experiment with the tail index values of $\xi_U= 0.19$, $0.39$, $0.59$, $0.79$, $0.99$, $1.19$, $1.39$, $1.59$, $1.79$ and $1.99$.
Note that $\xi_U \in \{0.19, 0.39, 0.59, 0.79, 0.99\}$ satisfy the condition for the consistency, but $\xi_U \in \{1.19, 1.39, 1.59, 1.79, 1.99\}$ fail to satisfy it.
For testing the finite second moment condition for the asymptotic normality of $\sqrt{n}\left(\hat\theta - \theta_0\right)$, we experiment with the tail index values of $\xi_U= 0.09$, $0.19$, $0.29$, $0.39$, $0.49$, $0.59$, $0.69$, $0.79$, $0.89$ and $0.99$.
Note that $\xi_U \in \{0.09, 0.19, 0.29, 0.39, 0.49\}$ satisfy the condition for the asymptotic normality, but $\xi_U \in \{0.59, 0.69, 0.79, 0.89, 0.99\}$ fail to satisfy it.
We also experiment with various numbers $k= 50$, $100$, and $200$ of order statistics for construction of the test.
Each set of simulations consists of 5000 Monte Carlo iterations.

%%%%%%%%%%%%%%%%%%%%%%%%%%%%%%%%%%%%%%%%%%%%%%%%%%%%%%%%%%%%%%%%%%%%%
\subsection{Simulation Results}
%%%%%%%%%%%%%%%%%%%%%%%%%%%%%%%%%%%%%%%%%%%%%%%%%%%%%%%%%%%%%%%%%%%%%

%%%%%%%%%%%%%%%%%%%%%%%%%%%%%%%%%%%%%%%%%%%%%%%%%%%%%%%%%%%%%%%%%%%%%
\begin{table}
	\centering
	\scalebox{0.85}{
	\begin{tabular}{crrrrrrrrrrrr}
	\multicolumn{13}{c}{(A) Linear Regression Model}\\
		\hline\hline
	 &&  \multicolumn{3}{c}{$n=10^4$} && \multicolumn{3}{c}{$n=10^5$} && \multicolumn{3}{c}{$n=10^6$} \\  
	\cline{3-5}\cline{7-9}\cline{11-13}
	$\xi_U$ && $k=$50 & $k=$100 & $k=$200 && $k=$50 & $k=$100 & $k=$200 && $k=$50 & $k=$100 & $k=$200 \\
	\hline
        0.19 && 0.00 & 0.00 & 0.00 && 0.00 & 0.00 & 0.00 && 0.00 & 0.00 & 0.00 \\
	0.39 && 0.00 & 0.00 & 0.00 && 0.00 & 0.00 & 0.00 && 0.00 & 0.00 & 0.00 \\
	0.59 && 0.00 & 0.00 & 0.00 && 0.00 & 0.00 & 0.00 && 0.00 & 0.00 & 0.00 \\
	0.79 && 0.01 & 0.00 & 0.00 && 0.01 & 0.00 & 0.00 && 0.01 & 0.00 & 0.00 \\
	0.99 && 0.06 & 0.07 & 0.09 && 0.05 & 0.06 & 0.06 && 0.05 & 0.05 & 0.05 \\
	\hline
	1.19 && 0.20 & 0.29 & 0.44 && 0.19 & 0.27 & 0.43 && 0.18 & 0.25 & 0.43 \\
	1.39 && 0.40 & 0.58 & 0.71 && 0.39 & 0.60 & 0.81 && 0.40 & 0.60 & 0.83 \\
	1.59 && 0.57 & 0.72 & 0.69 && 0.60 & 0.81 & 0.91 && 0.62 & 0.83 & 0.96 \\
	1.79 && 0.71 & 0.74 & 0.60 && 0.76 & 0.89 & 0.88 && 0.77 & 0.94 & 0.97 \\
	1.99 && 0.76 & 0.70 & 0.51 && 0.86 & 0.90 & 0.82 && 0.88 & 0.97 & 0.95 \\
		\hline\hline
		\\
	\multicolumn{13}{c}{(B) Linear IV Model}\\
		\hline\hline
	 &&  \multicolumn{3}{c}{$n=10^4$} && \multicolumn{3}{c}{$n=10^5$} && \multicolumn{3}{c}{$n=10^6$} \\  
	\cline{3-5}\cline{7-9}\cline{11-13}
	$\xi_U$ && $k=$50 & $k=$100 & $k=$200 && $k=$50 & $k=$100 & $k=$200 && $k=$50 & $k=$100 & $k=$200 \\
	\hline
        0.19 && 0.00 & 0.00 & 0.00 && 0.00 & 0.00 & 0.00 && 0.00 & 0.00 & 0.00 \\
	0.39 && 0.00 & 0.00 & 0.00 && 0.00 & 0.00 & 0.00 && 0.00 & 0.00 & 0.00 \\
	0.59 && 0.00 & 0.00 & 0.00 && 0.00 & 0.00 & 0.00 && 0.00 & 0.00 & 0.00 \\
	0.79 && 0.01 & 0.00 & 0.00 && 0.01 & 0.00 & 0.00 && 0.01 & 0.00 & 0.00 \\
	0.99 && 0.06 & 0.07 & 0.09 && 0.06 & 0.05 & 0.04 && 0.04 & 0.05 & 0.06 \\
	\hline
	1.19 && 0.20 & 0.29 & 0.45 && 0.18 & 0.27 & 0.43 && 0.18 & 0.27 & 0.42 \\
	1.39 && 0.40 & 0.57 & 0.69 && 0.39 & 0.59 & 0.81 && 0.39 & 0.60 & 0.83 \\
	1.59 && 0.58 & 0.70 & 0.69 && 0.60 & 0.81 & 0.91 && 0.61 & 0.84 & 0.96 \\
	1.79 && 0.70 & 0.73 & 0.58 && 0.75 & 0.88 & 0.87 && 0.78 & 0.94 & 0.96 \\
	1.99 && 0.75 & 0.67 & 0.48 && 0.84 & 0.89 & 0.81 && 0.88 & 0.96 & 0.94 \\
		\hline\hline
		\end{tabular}
	}
	\caption{Rejection probabilities of the test (\ref{test}) of the consistency of $\hat\theta$ in (A) the linear regression model and (B) the linear IV model. The results are based on 5000 simulation draws. The significance level is 0.05.}
	\label{tab:mc_first}
\end{table}
%%%%%%%%%%%%%%%%%%%%%%%%%%%%%%%%%%%%%%%%%%%%%%%%%%%%%%%%%%%%%%%%%%%%%

%%%%%%%%%%%%%%%%%%%%%%%%%%%%%%%%%%%%%%%%%%%%%%%%%%%%%%%%%%%%%%%%%%%%%
\begin{table}
	\centering
	\scalebox{0.85}{
	\begin{tabular}{crrrrrrrrrrrr}
	\multicolumn{13}{c}{(A) Linear Regression Model}\\
		\hline\hline
	 &&  \multicolumn{3}{c}{$n=10^4$} && \multicolumn{3}{c}{$n=10^5$} && \multicolumn{3}{c}{$n=10^6$} \\  
	\cline{3-5}\cline{7-9}\cline{11-13}
	$\xi_U$ && $k=$50 & $k=$100 & $k=$200 && $k=$50 & $k=$100 & $k=$200 && $k=$50 & $k=$100 & $k=$200 \\
	\hline
	0.09 && 0.00 & 0.00 & 0.00 && 0.00 & 0.00 & 0.00 && 0.00 & 0.00 & 0.00 \\
	0.19 && 0.00 & 0.00 & 0.00 && 0.00 & 0.00 & 0.00 && 0.00 & 0.00 & 0.00 \\
	0.29 && 0.00 & 0.00 & 0.00 && 0.00 & 0.00 & 0.00 && 0.00 & 0.00 & 0.00 \\
	0.39 && 0.01 & 0.01 & 0.00 && 0.01 & 0.00 & 0.00 && 0.01 & 0.00 & 0.00 \\
	0.49 && 0.07 & 0.05 & 0.06 && 0.07 & 0.05 & 0.04 && 0.06 & 0.05 & 0.03 \\
	\hline
	0.59 && 0.19 & 0.27 & 0.47 && 0.18 & 0.25 & 0.41 && 0.17 & 0.24 & 0.39 \\
	0.69 && 0.44 & 0.59 & 0.81 && 0.43 & 0.59 & 0.78 && 0.43 & 0.58 & 0.77 \\
	0.79 && 0.60 & 0.85 & 0.98 && 0.60 & 0.84 & 0.98 && 0.60 & 0.83 & 0.98 \\
	0.89 && 0.81 & 0.95 & 0.99 && 0.81 & 0.95 & 1.00 && 0.80 & 0.95 & 1.00 \\
	0.99 && 0.87 & 0.98 & 0.99 && 0.88 & 0.99 & 1.00 && 0.88 & 0.99 & 1.00 \\
		\hline\hline
		\\
	\multicolumn{13}{c}{(B) Linear IV Model}\\
		\hline\hline
	 &&  \multicolumn{3}{c}{$n=10^4$} && \multicolumn{3}{c}{$n=10^5$} && \multicolumn{3}{c}{$n=10^6$} \\  
	\cline{3-5}\cline{7-9}\cline{11-13}
	$\xi_U$ && $k=$50 & $k=$100 & $k=$200 && $k=$50 & $k=$100 & $k=$200 && $k=$50 & $k=$100 & $k=$200 \\
	\hline
	0.09 && 0.00 & 0.00 & 0.00 && 0.00 & 0.00 & 0.00 && 0.00 & 0.00 & 0.00 \\
	0.19 && 0.00 & 0.00 & 0.00 && 0.00 & 0.00 & 0.00 && 0.00 & 0.00 & 0.00 \\
	0.29 && 0.00 & 0.00 & 0.00 && 0.00 & 0.00 & 0.00 && 0.00 & 0.00 & 0.00 \\
	0.39 && 0.01 & 0.01 & 0.00 && 0.01 & 0.00 & 0.00 && 0.00 & 0.00 & 0.00 \\
	0.49 && 0.07 & 0.06 & 0.06 && 0.07 & 0.05 & 0.04 && 0.06 & 0.05 & 0.03 \\
	\hline
	0.59 && 0.18 & 0.29 & 0.48 && 0.18 & 0.25 & 0.41 && 0.18 & 0.25 & 0.39 \\
	0.69 && 0.44 & 0.61 & 0.81 && 0.43 & 0.60 & 0.79 && 0.43 & 0.57 & 0.77 \\
	0.79 && 0.61 & 0.83 & 0.98 && 0.59 & 0.83 & 0.98 && 0.60 & 0.82 & 0.98 \\
	0.89 && 0.81 & 0.95 & 0.99 && 0.81 & 0.95 & 1.00 && 0.80 & 0.95 & 1.00 \\
	0.99 && 0.87 & 0.98 & 0.98 && 0.88 & 0.99 & 1.00 && 0.88 & 0.99 & 1.00 \\
		\hline\hline
		\end{tabular}
	}
	\caption{Rejection probabilities of the test (\ref{test}) of the asymptotic normality of $\sqrt{n}\left(\hat\theta - \theta_0\right)$ in (A) the linear regression model and (B) the linear IV model. The results are based on 5000 simulation draws. The significance level is 0.05.}
	${}$\\
	\label{tab:mc_second}
\end{table}
%%%%%%%%%%%%%%%%%%%%%%%%%%%%%%%%%%%%%%%%%%%%%%%%%%%%%%%%%%%%%%%%%%%%%

Table \ref{tab:mc_first} shows Monte Carlo simulation results of testing the finite first moment condition for the consistency of $\hat\theta$ in (A) the linear regression model and (B) the linear IV model.
In both of the two panels, (A) and (B), we can see that the simulated rejection probabilities are dominated by the nominal size 0.05 for all of $\xi_U \in \{0.19, 0.39, 0.59, 0.79\}$ in the null region, and those are approximately the same as the nominal size 0.05 near the boundary, i.e., $\xi_U = 0.99$, of the null region.
These results support the uniform size control property of the test that is established in Theorem \ref{theorem:main} as the main result of this paper.
The condition for the consistency holds for any of $\xi_U < 1$, but a researcher does not \textit{ex ante} know or does not want to fix which exact value $\xi_U$ takes for a specific application under the composite null hypothesis in consideration.
For this reason, this nearly uniform size control property is important in practice.

Table \ref{tab:mc_second} shows Monte Carlo simulation results of testing the finite second moment condition for the asymptotic normality of $\sqrt{n}\left(\hat\theta - \theta_0\right)$ in (A) the linear regression model and (B) the linear IV model. 
The findings here are very similar to those in the consistency test presented above.
Namely, in both of the two panels, (A) and (B), we can see that the simulated rejection probabilities are dominated by the nominal size 0.05 for all of $\xi_U \in \{0.09, 0.19, 0.29, 0.39\}$ in the null region, and those are approximately the same as the nominal size 0.05 near the boundary, i.e., $\xi_U = 0.49$, of the null region.
Again, these results support the nearly uniform size control property of the test that is established in Theorem \ref{theorem:main}.

%%%%%%%%%%%%%%%%%%%%%%%%%%%%%%%%%%%%%%%%%%%%%%%%%%%%%%%%%%%%%%%%%%%%%
\section{Application to Demand Estimation}\label{sec:application}
%%%%%%%%%%%%%%%%%%%%%%%%%%%%%%%%%%%%%%%%%%%%%%%%%%%%%%%%%%%%%%%%%%%%%

In this section, we present an empirical application of the proposed test procedure.
Recall the framework of demand estimation in differentiated products markets introduced in Example \ref{example:blp}.
The dependent variable is defined by the logarithm of the market share of a product relative to that of an outside product.
In rich data sets, we often encounter zero empirical market shares.
Since the logarithm of zero is undefined, empirical practitioners often use \textit{ad hoc} procedures to deal with observations with zero market share.
One common way is to simply remove observations with zero empirical market shares.
Another common way is to replace zeros with a small positive value.
Both of these two \textit{ad hoc} treatments result in biased estimates in general, as demonstrated through Monte Carlo simulation studies by \citet*{GandhiLuShi2017}.
In implementing the second approach, empirical researchers often substitute infinitesimal positive values $\Delta$ for zeros, perhaps in efforts to mitigate such biases.
In this paper, we show that substitution of infinitesimal positive values $\Delta$ in fact results in pathetic asymptotic behaviors of the estimator.
Specifically, such an \textit{ad hoc} estimator fails the root-n asymptotic normality, as we reject the finite second moment condition of the score.
Furthermore, such an estimator is not even likely to converge in probability to a possibly biased pseudo-true target either, as we reject the finite first moment condition of the score too.
These results follow because the introduction of the a huge negative number (as the logarithm of an infinitesimal number) turns some of the observations with \textit{originally non-zero} shares into outliers, as we will carefully illustrate ahead after presenting the test results.

Following preceding papers on market analysis, we use scanner data from the Dominick's Finer Foods (DFF) retail chain.\footnote{We thank James M. Kilts Center, University of Chicago Booth School of Business for allowing us to use this data set. It is available at https://www.chicagobooth.edu/research/kilts/datasets/dominicks.}
The unit of observation is defined by the product of UPC (universal product code), store, and week.
Our analysis, as described below, follows that of \citet*{GandhiLuShi2017}.
We focus on the product category of canned tuna.
Empirical market shares are constructed by using quantity sales and the number of customers who visited the store in the week.
Control variables include the price, UPC fixed effects, and a time trend.
We instrument the possibly endogenous prices by the wholesale costs, which are calculated by inverting the gross margin.

The number of observations is approximately $10^6$, similar to the sample sizes considered in our Monte Carlo simulation studies in Section \ref{sec:simulation}. 
This feature of the data allows us to use a reasonably large number $k$ of order statistics to enhance the power of our proposed test.
Among this large number of observations, approximately 44\% of the observations are recorded to have zero empirical market share.
The smallest non-zero empirical market share is approximately $10^{-5}$.
Therefore, it is sensible to replace the zero empirical market share by an infinitesimal positive number $\Delta$ that is no larger than $10^{-5}$.
In our analysis, therefore, we consider the following numbers to replace zero: $\Delta=$ $10^{-5}$, $10^{-6}$, ..., $10^{-19}$, $10^{-20}$.

Table \ref{tab:application_tuna_first} summarizes the p-values of testing the finite first moment condition for the consistency.
Similarly, Table \ref{tab:application_tuna_second} summarizes the p-values of testing the finite second moment condition for the root-n asymptotic normality.
For the sake of transparency, we show results for various numbers of $k$ ranging from 1000 to 5000.
Before discussing these results, first note that small numbers $k$ of order statistics in general entail short power.
In view of Figure \ref{fig:power_curves}, we can see that $k$ ranging from 1000 to 5000 yields very strong powers of the test.
Furthermore, note also that the number $k=1000$ corresponds to only 0.1 percent of the whole sample, so that the extreme value approximation should perform well.
With these in mind, observe that the results reported in Tables \ref{tab:application_tuna_first} and \ref{tab:application_tuna_second} suggest that we start to reject the null hypothesis of a finite first and second moments when $k$ is larger than 1000.
The rejection of the finite second moment conditions (Table \ref{tab:application_tuna_second}) implies that the root-n asymptotic normality of the demand estimator may perform poorly if we conduct the \textit{ad hoc} practice of replacing the zero empirical market share by any of the infinitesimal positive values $\Delta=$ $10^{-5}$, $10^{-6}$, ..., $10^{-19}$, $10^{-20}$.
Furthermore, the rejection of the finite first moment conditions (Table \ref{tab:application_tuna_first}) implies that such an \textit{ad hoc} estimator may not even converge in probability to a possibly biased pseudo-true target.

%%%%%%%%%%%%%%%%%%%%%%%%%%%%%%%%%%%%%%%%%%%%%%%%%%%%%%%%%%%%%%%%%%%%%
\begin{table}
	\centering
	\begin{tabular}{ccccccccccc}
		\\\\\hline\hline
	$\Delta$ && $k=$1000 & $k$=2000 & $k=$3000 & $k=$4000 & $k=$5000 \\
	\hline
         $10^{-5}$ && 0.67 & 0.00 & 0.00 & 1.00 & 1.00 \\
	 $10^{-6}$ && 1.00 & 0.00 & 0.00 & 0.00 & 0.23 \\
	 $10^{-7}$ && 1.00 & 0.00 & 0.00 & 0.00 & 0.00 \\
	 $10^{-8}$ && 1.00 & 0.00 & 0.00 & 0.00 & 0.00 \\
	 $10^{-9}$ && 1.00 & 0.00 & 0.00 & 0.00 & 0.00\\
	$10^{-10}$ && 1.00 & 0.00 & 0.00 & 0.00 & 0.00 \\
	$10^{-11}$ && 1.00 & 0.00 & 0.00 & 0.00 & 0.00  \\
	$10^{-12}$ && 1.00 & 0.00 & 0.00 & 0.00 & 0.00 \\
	$10^{-13}$ && 1.00 & 0.00 & 0.00 & 0.00 & 0.00 \\
	$10^{-14}$ && 1.00 & 0.00 & 0.00 & 0.00 & 0.00 \\
	$10^{-15}$ && 1.00 & 0.00 & 0.00 & 0.00 & 0.00 \\
	$10^{-16}$ && 1.00 & 0.00 & 0.00 & 0.00 & 0.00 \\
	$10^{-17}$ && 1.00 & 0.00 & 0.00 & 0.00 & 0.00 \\
	$10^{-18}$ && 1.00 & 0.00 & 0.00 & 0.00 & 0.00 \\
	$10^{-19}$ && 1.00 & 0.00 & 0.00 & 0.00 & 0.00 \\
	$10^{-20}$ && 1.00 & 0.00 & 0.00 & 0.00 & 0.00 \\
		\hline\hline
		\end{tabular}
	\caption{P-values of the test (\ref{test}) of the finite first moment condition for consistency with the market share data from DFF for the product category of canned tuna data, where zero empirical market shares are replaced by $\Delta=$ $10^{-5}$, $10^{-6}$, ..., $10^{-19}$, $10^{-20}$.}
	${}$\\${}$
	\label{tab:application_tuna_first}
\end{table}
%%%%%%%%%%%%%%%%%%%%%%%%%%%%%%%%%%%%%%%%%%%%%%%%%%%%%%%%%%%%%%%%%%%%%

%%%%%%%%%%%%%%%%%%%%%%%%%%%%%%%%%%%%%%%%%%%%%%%%%%%%%%%%%%%%%%%%%%%%%
\begin{table}
	\centering
	\begin{tabular}{ccccccc}
		\\\\\hline\hline
	$\Delta$ && $k=$1000 & $k$=2000 & $k=$3000 & $k=$4000 & $k=$5000 \\
	\hline
         $10^{-5}$ && 0.00 & 0.00 & 0.00 & 0.00 & 0.00 \\
	 $10^{-6}$ && 0.00 & 0.00 & 0.00 & 0.00 & 0.00 \\
	 $10^{-7}$ && 0.00 & 0.00 & 0.00 & 0.00 & 0.00 \\
	 $10^{-8}$ && 0.01 & 0.00 & 0.00 & 0.00 & 0.00 \\
	 $10^{-9}$ && 0.02 & 0.00 & 0.00 & 0.00 & 0.00 \\
	$10^{-10}$ && 0.04 & 0.00 & 0.00 & 0.00 & 0.00 \\
	$10^{-11}$ && 0.06 & 0.00 & 0.00 & 0.00 & 0.00 \\
	$10^{-12}$ && 0.08 & 0.00 & 0.00 & 0.00 & 0.00 \\
	$10^{-13}$ && 0.10 & 0.00 & 0.00 & 0.00 & 0.00 \\
	$10^{-14}$ && 0.13 & 0.00 & 0.00 & 0.00 & 0.00 \\
	$10^{-15}$ && 0.14 & 0.00 & 0.00 & 0.00 & 0.00 \\
	$10^{-16}$ && 0.16 & 0.00 & 0.00 & 0.00 & 0.00 \\
	$10^{-17}$ && 0.18 & 0.00 & 0.00 & 0.00 & 0.00 \\
	$10^{-18}$ && 0.21 & 0.00 & 0.00 & 0.00 & 0.00 \\
	$10^{-19}$ && 0.22 & 0.00 & 0.00 & 0.00 & 0.00 \\
	$10^{-20}$ && 0.23 & 0.00 & 0.00 & 0.00 & 0.00 \\
		\hline\hline
		\end{tabular}
	\caption{P-values of the test (\ref{test}) of the finite second moment condition for the root-n asymptotic normality with the market share data from DFF for the product category of canned tuna data, where zero empirical market shares are replaced by $\Delta=$ $10^{-5}$, $10^{-6}$, ..., $10^{-19}$, $10^{-20}$.}
	${}$\\${}$
	\label{tab:application_tuna_second}
\end{table}
%%%%%%%%%%%%%%%%%%%%%%%%%%%%%%%%%%%%%%%%%%%%%%%%%%%%%%%%%%%%%%%%%%%%%

\begin{figure}
	\centering
	\begin{tabular}{cc}
		(A) $\log$(share) on $A^1(\hat\theta)$ for $\Delta=10^{-5}$ & 
		(B) $\log$(share) on $A^2(\hat\theta)$ for $\Delta=10^{-5}$\\
		\includegraphics[width=0.5\textwidth]{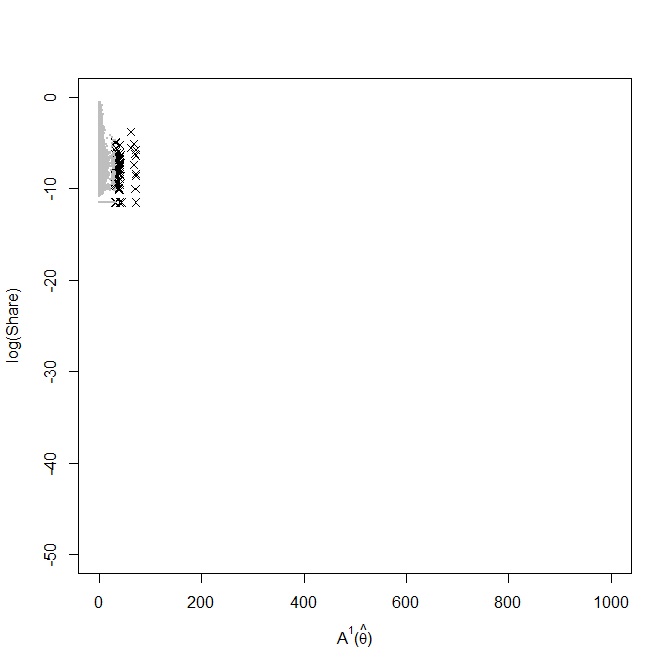} &
		\includegraphics[width=0.5\textwidth]{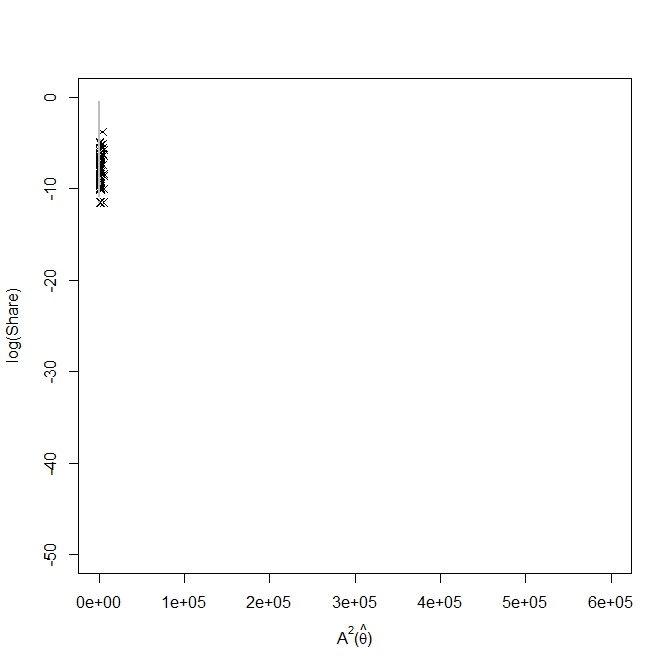} \\
		(C) $\log$(share) on $A^1(\hat\theta)$ for $\Delta=10^{-10}$ & 
		(D) $\log$(share) on $A^2(\hat\theta)$ for $\Delta=10^{-10}$\\
		\includegraphics[width=0.5\textwidth]{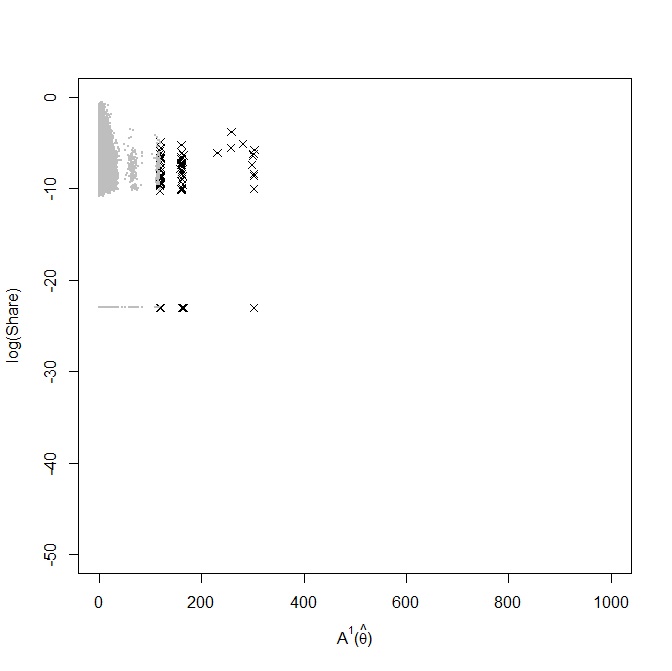} &
		\includegraphics[width=0.5\textwidth]{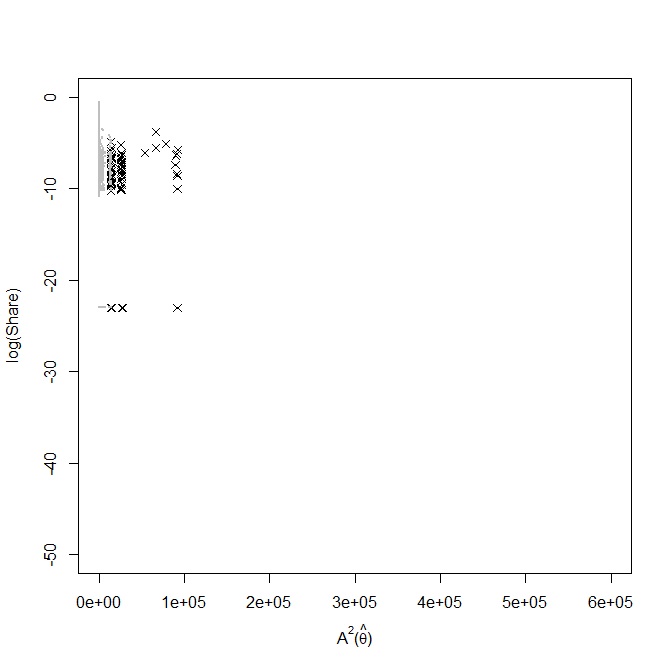} \\
	\end{tabular}
	\caption{Scatter plots of (A) $\log$(share) on $A^1(\hat\theta)$ for $\Delta=10^{-5}$; (B) $\log$(share) on $A^2(\hat\theta)$ for $\Delta=10^{-5}$; (C) $\log$(share) on $A^1(\hat\theta)$ for $\Delta=10^{-10}$; and (D) $\log$(share) on $A^2(\hat\theta)$ for $\Delta=10^{-10}$. Observations with the top 0.0001-quantile of $A^1(\hat\theta)$ and $A^2(\hat\theta)$ are marked by black crosses.}
	\label{fig:scatter_plots1}
\end{figure}

\begin{figure}
	\centering
	\begin{tabular}{cc}
		(A) $\log$(share) on $A^1(\hat\theta)$ for $\Delta=10^{-15}$ & 
		(B) $\log$(share) on $A^2(\hat\theta)$ for $\Delta=10^{-15}$\\
		\includegraphics[width=0.5\textwidth]{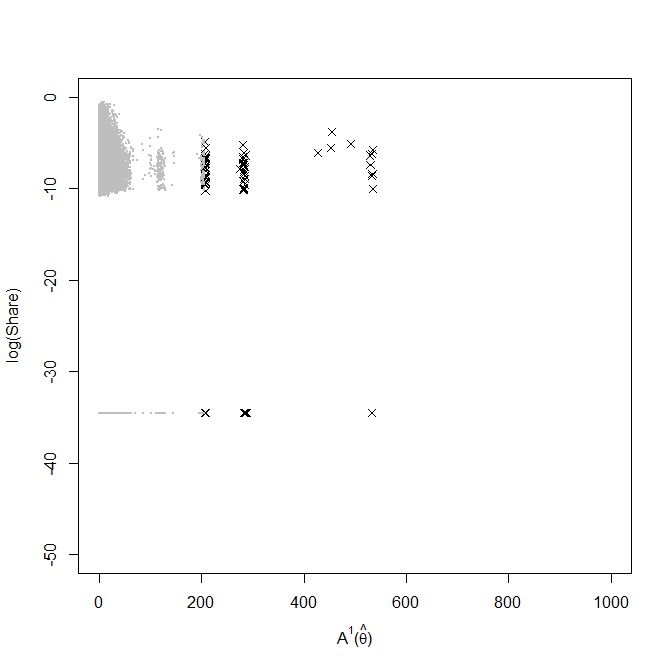} &
		\includegraphics[width=0.5\textwidth]{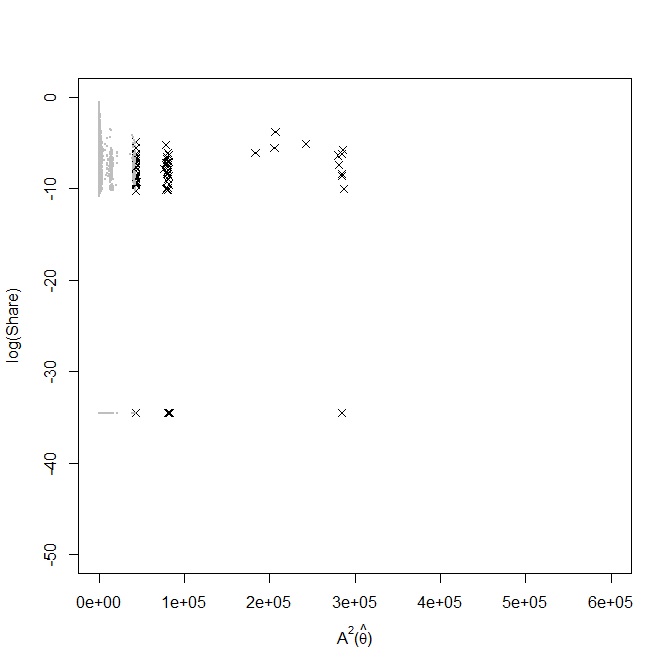} \\
		(C) $\log$(share) on $A^1(\hat\theta)$ for $\Delta=10^{-20}$ & 
		(D) $\log$(share) on $A^2(\hat\theta)$ for $\Delta=10^{-20}$\\
		\includegraphics[width=0.5\textwidth]{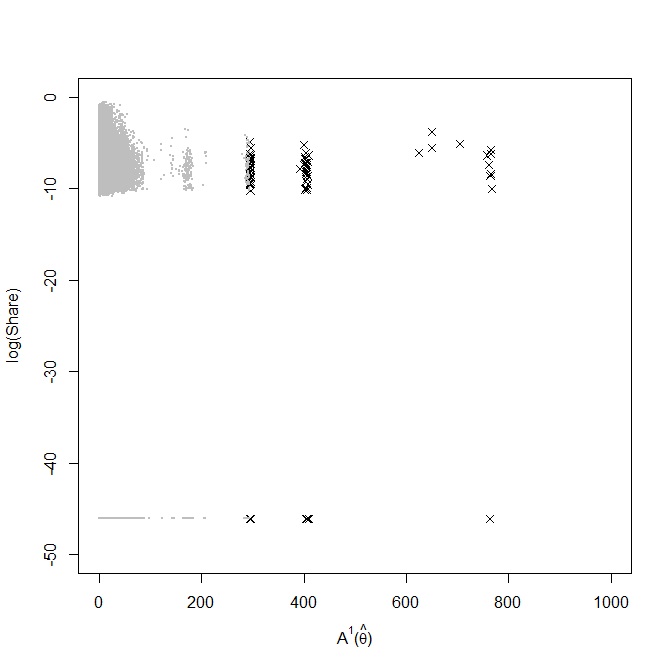} &
		\includegraphics[width=0.5\textwidth]{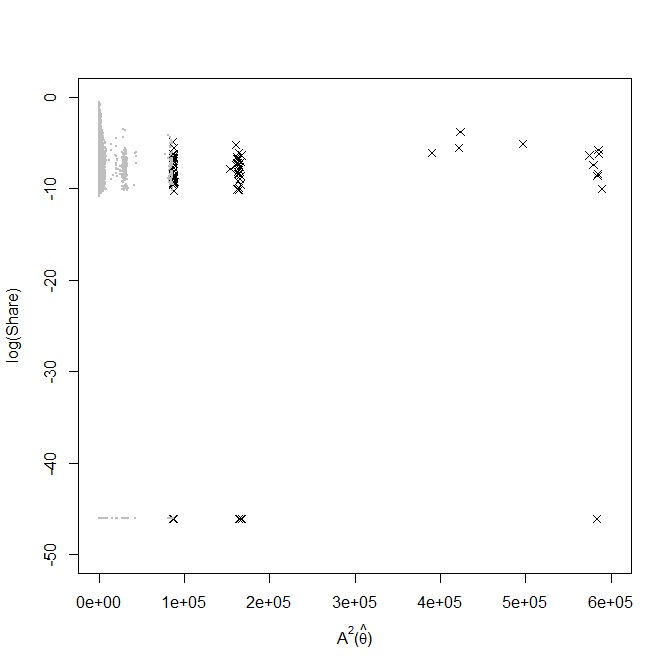} \\
	\end{tabular}
	\caption{Scatter plots of (A) $\log$(share) on $A^1(\hat\theta)$ for $\Delta=10^{-15}$; (B) $\log$(share) on $A^2(\hat\theta)$ for $\Delta=10^{-15}$; (C) $\log$(share) on $A^1(\hat\theta)$ for $\Delta=10^{-20}$; and (D) $\log$(share) on $A^2(\hat\theta)$ for $\Delta=10^{-20}$. Observations with the top 0.0001-quantile of $A^1(\hat\theta)$ and $A^2(\hat\theta)$ are marked by black crosses.}
	\label{fig:scatter_plots2}
\end{figure}

While the test rejects the null hypotheses of finite moments of $A_i^1(\theta_0)$ and $A_i^2(\theta_0)$, a natural question is why the \textit{ad hoc} procedure of adding a small constant to the zero market share causes the heavy tailed distributions of $A_i^1(\theta_0)$ and $A_i^2(\theta_0)$.
Since the logarithm of a small constant is finite anyway, it appears to only produce a 44\% point mass of absolutely very large yet finite constants.
As such, these small numbers do not seem to contribute to heavy tails \textit{by themselves}. 
To see what is going on behind our test rejecting the null hypotheses, we display eight scatter plots in Figures \ref{fig:scatter_plots1} and \ref{fig:scatter_plots2}.
Figure \ref{fig:scatter_plots1} displays plots of (A) $\log$(share) on $A^1(\hat\theta)$ for $\Delta=10^{-5}$; (B) $\log$(share) on $A^2(\hat\theta)$ for $\Delta=10^{-5}$; (C) $\log$(share) on $A^1(\hat\theta)$ for $\Delta=10^{-10}$; and (D) $\log$(share) on $A^2(\hat\theta)$ for $\Delta=10^{-10}$.
Figure \ref{fig:scatter_plots2} displays plots of (A) $\log$(share) on $A^1(\hat\theta)$ for $\Delta=10^{-15}$; (B) $\log$(share) on $A^2(\hat\theta)$ for $\Delta=10^{-15}$; (C) $\log$(share) on $A^1(\hat\theta)$ for $\Delta=10^{-20}$; and (D) $\log$(share) on $A^2(\hat\theta)$ for $\Delta=10^{-20}$.
Those observations above the top 0.0001-quantile of $A^1(\hat\theta)$ and $A^2(\hat\theta)$ are marked by black crosses, while all else are marked by gray dots.

%%%%%%%%%%%%%%%%%%%%%%%%%%%%%%%%%%%%%%%%%%%%%%%%%%%%%%%%%%%%%%%%%%%%%
%%%%%%%%%%%%%%%%%%%%%%%%%%%%%%%%%%%%%%%%%%%%%%%%%%%%%%%%%%%%%%%%%%%%%

%%%%%%%%%%%%%%%%%%%%%%%%%%%%%%%%%%%%%%%%%%%%%%%%%%%%%%%%%%%%%%%%%%%%%
%%%%%%%%%%%%%%%%%%%%%%%%%%%%%%%%%%%%%%%%%%%%%%%%%%%%%%%%%%%%%%%%%%%%%

In each of the panels in Figures \ref{fig:scatter_plots1} and \ref{fig:scatter_plots2}, note that the observations with originally zero market share appear on the horizontal line at the vertical level of $\log(\Delta)$.
As $\Delta$ becomes smaller, these lines move downward and they tend to behave as observations with an absolutely large $Y$ value.
However, these observations with originally zero shares are not necessarily outliers by themselves because as many as 44\% of the observations exist on this line.
Instead, many of the outliers (i.e., observations marked by the black crosses) stem from the group of observations with \textit{originally non-zero} market shares.
Furthermore, the \textit{horizontal} distances between those marked by the black crosses and the major cluster of observations marked by gray dots widen as $\Delta$ becomes smaller, i.e., the \textit{horizontal} spread is the smallest in Figure \ref{fig:scatter_plots1} (A)--(B) and the largest in Figure \ref{fig:scatter_plots2} (C)--(D).
This pattern implies that, while most of the 44\% of observations with originally zero market share are not outliers by themselves despite the isolated levels of $\log(\Delta)$, smaller values of $\Delta$ are turning some of the observations with originally non-zero market shares into outliers to larger extents.

We conclude this section by discussing the implications of our test results and practical suggestions in light of them.
Rich market share data often include zero empirical market shares.
Since the logarithm of zero is undefined, empirical researchers often employ the \textit{ad hoc} practice of replacing the zero by an infinitesimal positive value $\Delta$.
This practice has already been known to incur biased estimates \citep*[see][]{GandhiLuShi2017}, but can also result in a failure of the root-n asymptotic normality in addition.
Furthermore, the \textit{ad hoc} estimator is not even guaranteed to converge in probability to a possibly biased pseudo-true target either.
As such, both the point estimates and their standard errors are incredible.
Empirical researchers may, therefore, want to resort to alternative methods that are robust against zero market shares, such as the method proposed by \citet*{GandhiLuShi2017}.

%%%%%%%%%%%%%%%%%%%%%%%%%%%%%%%%%%%%%%%%%%%%%%%%%%%%%%%%%%%%%%%%%%%%%
\section{Summary and Discussions}\label{sec:summary}
%%%%%%%%%%%%%%%%%%%%%%%%%%%%%%%%%%%%%%%%%%%%%%%%%%%%%%%%%%%%%%%%%%%%%

Many empirical studies in economics rely on the GMM and M estimators including, but not limited to, the OLS, GLS, QMLE, and 2SLS.
Furthermore, they usually rely on the consistency and the root-n asymptotic normality of these estimators when drawing scientific conclusions via statistical inference.
Although the conditions for the consistency and the root-n asymptotic normality are usually taken for granted as such, they may not be always plausibly satisfied.
In this light, this paper proposes a method of testing the hypothesis of finite first and second moments of scores, which serve as key conditions of the consistency and the root-n asymptotic normality, respectively.

There are two desired properties of our proposed test in practice.
First, unlike other approaches in extreme value theory that require a sequence of tuning parameter values that change as the sample size grows, our test is valid for any predetermined fixed number $k$ of order statistics to be used to construct the test.
This is a useful property in practice because it relieves researchers from worrying about a `valid' data driven choice of tuning parameters for the purpose of size control.
Second, our test has a nearly uniform size control property over the set of data generating processes for which the asymptotic normality holds.
This nearly uniform size control property is useful in practice, because researchers usually do not \textit{ex ante} know or do not want to fix the true tail index $\xi$ in the applications of their interest under the composite null hypothesis in consideration.
Monte Carlo simulation studies indeed support this theoretical property for two of the most commonly used econometric frameworks, namely the linear regression model and the linear IV model.

A failure of the consistency and the root-n asymptotic normality may be caused by the following two cases among others.
First, some dependent variables (e.g., wealth, infant birth weight, murder rate) are reported to exhibit heavy tailed distributions, and they can induce infinite first and second moments of the score of an estimator.
Second, when a dependent variable is the logarithm of a variable, practitioners sometimes employ an \textit{ad hoc} procedure of replacing zeros by infinitesimal values.
This practice can lead to heavy tailed distribution of the residuals for observations with originally non-zero market shares.
In our empirical application, we highlighted the latter case.
Using scanner data from the Dominick's Finer Foods (DFF) retail chain, we reject the consistency and the root-n asymptotic normality for demand estimators based on such an \textit{ad hoc} practice.

Finally, we conclude this paper by remarking that the test can be used to enhance the quality and credibility of past and future empirical studies.
On one hand, if our test supports the finite moment conditions for consistency and the root-n asymptotic normality for a selected empirical work, then the test result reinforces the credibility of scientific conclusions reported by that work.
On the other hand, if our test fails to support the finite moment conditions for a selected empirical work, then a researcher may want to consider one of the alternative robust approaches for more credible empirical research.

%%%%%%%%%%%%%%%%%%%%%%%%%%%%%%%%%%%%%%%%%%%%%%%%%%%%%%%%%%%%%%%%%%%%%
\appendix
\section*{Appendix}
%%%%%%%%%%%%%%%%%%%%%%%%%%%%%%%%%%%%%%%%%%%%%%%%%%%%%%%%%%%%%%%%%%%%%

The appendix consists of two sections.
Appendix \ref{sec:proofs} contains proofs of the main results, namely Lemma \ref{lemma:ev} and Theorem \ref{theorem:main}.
Appendix \ref{sec:computational_algorithm} contains additional computational details to implement the test.

%%%%%%%%%%%%%%%%%%%%%%%%%%%%%%%%%%%%%%%%%%%%%%%%%%%%%%%%%%%%%%%%%%%%%
\section{Proofs}\label{sec:proofs}
%%%%%%%%%%%%%%%%%%%%%%%%%%%%%%%%%%%%%%%%%%%%%%%%%%%%%%%%%%%%%%%%%%%%%

%%%%%%%%%%%%%%%%%%%%%%%%%%%%%%%%%%%%%%%%%%%%%%%%%%%%%%%%%%%%%%%%%%%%%
\subsection{Proof of Lemma \ref{lemma:ev}}\label{sec:lemma:ev}
%%%%%%%%%%%%%%%%%%%%%%%%%%%%%%%%%%%%%%%%%%%%%%%%%%%%%%%%%%%%%%%%%%%%%
\paragraph*{Proof of Lemma 1}
First, by the EV theory, Condition \ref{conditions}.(i) ($D _{i}$ is i.i.d.) and
Condition \ref{conditions}.(ii) ($F_{A^r \left(\theta _{0}\right)}\in \mathcal{D}\left( G_{\xi} \right) $) imply 
\begin{equation}
 \frac{\mathbf{A}^r\left( \theta _{0}\right) -b_{n}}{a_{n}}
\overset{d}{\rightarrow }\mathbf{V},  \label{EVT1}
\end{equation}%
where $\mathbf{V}$ is jointly EV distributed with tail index $\xi$. By Corollary 1.2.4 and Remark 1.2.7 in \citet*{HaanFerreira2006},
the constants $a_{n}$ and $b_{n}$ can be chosen as follows. If $\xi>0$, we choose $a_{n}=Q_{A^r\left( \theta _{0}\right) }\left(
1-1/n\right) $ and $b_{n}=0$. If $\xi=0$, we choose $%
a_{n}=1/\left( nf_{A^r\left( \theta _{0}\right) }\left( b_{n}\right) \right) $
and $b_{n}=Q_{A^r\left( \theta _{0}\right) }\left( 1-1/n\right) $. By
construction, these constants satisfy that $1-F_{A^r \left(\theta _{0}\right)}\left(
a_{n}y+b_{n}\right) =O\left( n^{-1}\right) $ for every $y>0$ in both of the cases.

Now, let $I=(I_{1},\ldots ,I_{k})\in \{1,\ldots ,n\}^{k}$ be the $k$ random
indices such that $A_{\left( j\right) }^r\left( \theta _{0}\right)
=A_{I_{j}}^r\left( \theta _{0}\right) $, $j=1,\ldots ,k$, and let $\hat{I}$ be
the corresponding indices such that $A_{\left( j\right) }^r\left( \hat{\theta}%
\right) =A_{\hat{I}_{j}}^r\left( \hat{\theta}\right) $. Then, the convergence
of $\mathbf{A}^r\left( \hat{\theta}\right) $ follows from (\ref{EVT1}) once we
establish $|A_{\hat{I}_{j}}^r(\hat{\theta})-A_{I_{j}}^r\left( \theta _{0}\right)
|=o_{p}(a_{n})$ for $j=1,\ldots ,k$. We present the case of $k=1$, but the argument for a general $k$ is similar. Denote $\varepsilon _{i}\equiv A_{i}^r(\hat{\theta})-A_{i}^r\left( \theta _{0}\right) $.

First, consider the case with $\xi>0$. The part of Condition \ref{conditions}.(iii) for the case of $\xi \left. >\right. 0$ yields that 
\begin{eqnarray*}
\sup_{i}\left\vert \varepsilon _{i}\right\vert &=&\sup_{i}\left\vert A_{i}^r(%
\hat{\theta})-A_{i}^r\left( \theta _{0}\right) \right\vert \\
&\leq &\sup_{i}\sup_{\theta \in B_{\eta_{n} }\left( \theta _{0}\right)
}\left\vert \left\vert \frac{\partial A_{i}^r\left( \theta \right) }{\partial
\theta }\right\vert \right\vert \left\vert \left\vert \hat{\theta}-\theta
_{0}\right\vert \right\vert \\
&=&o_{p}(a_{n}).
\end{eqnarray*}%
Given this result, we have that, on one hand, $A_{\hat{I}}^r(\hat{\theta}%
)=\max_{i}\{A_{i}^r\left( \theta _{0}\right) +\varepsilon _{i}\}\leq
A_{I}^r\left( \theta _{0}\right) +\sup_{i}\left\vert \varepsilon
_{i}\right\vert =A_{I}^r\left( \theta _{0}\right) +o_{p}(a_{n})$; and, on the
other hand, $A_{\hat{I}}^r(\hat{\theta})=\max_{i}\{A_{i}^r\left( \theta
_{0}\right) +\varepsilon _{i}\}\geq \max_{i}\{A_{i}^r\left( \theta _{0}\right)
+\min_{i}\{\varepsilon _{i}\}\}\geq A_{I}^r\left( \theta _{0}\right)
+\min_{i}\{\varepsilon _{i}\}\geq \alpha _{I}-\sup_{i}\left\vert \varepsilon
_{i}\right\vert =\alpha _{I}-o_{p}(a_{n})$. Therefore, $\left\vert A_{\hat{I}%
_{j}}^r(\hat{\theta})-A_{I_{j}}^r\left( \theta _{0}\right) \right\vert \leq
o_{p}(a_{n})$ holds.

Next, consider the case with $\xi=0$. The part of Condition \ref{conditions}.(iii) for the case of $\xi \left. =\right. 0$ implies that $\sup_{i}\sup_{\theta \in B_{\eta_{n} }\left( \theta _{0}\right) }\left\vert \left\vert \frac{\partial
A_{i}^r\left( \theta \right) }{\partial \theta }\right\vert \right\vert
=O_{p}\left( a_{n}\right) $. The rest of the proof is identical to the case with $\xi>0$. $\square$
%%%%%%%%%%%%%%%%%%%%%%%%%%%%%%%%%%%%%%%%%%%%%%%%%%%%%%%%%%%%%%%%%%%%%

%%%%%%%%%%%%%%%%%%%%%%%%%%%%%%%%%%%%%%%%%%%%%%%%%%%%%%%%%%%%%%%%%%%%%
\subsection{Proof of Theorem \ref{theorem:main}}\label{sec:theorem:main}
%%%%%%%%%%%%%%%%%%%%%%%%%%%%%%%%%%%%%%%%%%%%%%%%%%%%%%%%%%%%%%%%%%%%%
\paragraph*{Proof of Theorem 1}
By Lemma 1 and the continuous mapping theorem, we have $\mathbf{A}_{\ast }^r\left( 
\hat{\theta}\right) \overset{d}{\rightarrow }\mathbf{V}_{\ast }$. Write $%
\Lambda \left( \xi \right) =c\tilde{\Lambda}\left( \xi \right) $ for some
constant $c>0$ and $\tilde{\Lambda}\left( \xi \right) $ a positive measure
on $\left[ 0,1-\varepsilon \right] $. Since the density $f_{\mathbf{V}_{\ast
}}$ is continuous, $\mathbb{E}_{\xi }\left[ \varphi \left( \mathbf{V}_{\ast
}\right) \right] $ as a function of $\xi $ and $c$ is also continuous in
both arguments for any given $\tilde{\Lambda}\left( \cdot \right) $.
Therefore, we can choose a large enough $c$ so that $\sup_{\xi \in \left[
0,1-\varepsilon \right] }\mathbb{E}_{\xi }\left[ \varphi \left( \mathbf{V}%
_{\ast }\right) \right] \leq \alpha $. $\square$
%%%%%%%%%%%%%%%%%%%%%%%%%%%%%%%%%%%%%%%%%%%%%%%%%%%%%%%%%%%%%%%%%%%%%
\begin{remark}
{\small Since $\tilde{\Lambda}$ in the last part of the above proof can be
arbitrary in theory, we provide an empirical guide for determining a nearly
optimal $\tilde{\Lambda}$ in the following section. }
\end{remark}
%%%%%%%%%%%%%%%%%%%%%%%%%%%%%%%%%%%%%%%%%%%%%%%%%%%%%%%%%%%%%%%%%%%%%
\section{Additional Computational Details}\label{sec:computational_algorithm}
%%%%%%%%%%%%%%%%%%%%%%%%%%%%%%%%%%%%%%%%%%%%%%%%%%%%%%%%%%%%%%%%%%%%%

This section provides computational details about constructing the test (\ref{test}), which is based on the limiting observation $\mathbf{V}_\ast$. The density of $\mathbf{V}_\ast$ is given by (\ref{ev_pdf}), which is computed by Gaussian Quadrature. To construct the test (\ref{test}), we specify the weight $W$ to be the uniform distribution for simplicity of exposition. The weight $W$ reflects the importance attached by the econometrician to different alternatives, which can be easily changed. Then, it remains to determine a suitable candidate for the weight $\Lambda$. We do this by the generic algorithm provided by \citet*{ElliottMullerWatson2015}. 

To be specific, we use the same notation as in the proof of Theorem \ref{theorem:main} and consider $\Lambda =c\tilde{\Lambda}$, where $\tilde{\Lambda}$ is some
probability distribution function with support on $\Xi =[0,1-\varepsilon]$. Suppose that $\xi $ is randomly drawn from $\tilde{\Lambda}$ and $c$ satisfies that $\int P_{\xi}(\varphi (\mathbf{V}_{\ast })=1)d\tilde{\Lambda}(\xi)=\alpha $, where we slightly
abuse the notation so that the subscripts $\xi $ and $\Lambda $ emphasize that they determine the test. Denote the $W$-weighted average power as $P_{\Lambda }=\int_{(1-\varepsilon ,2)}P_{\xi}(\varphi _{\Lambda }(\mathbf{V}_{\ast })=1)dW(\xi )$. Since the uniform size constraint for all $\xi \in \Xi $ implies the $\tilde{\Lambda}$-weighted average size control for any probability distribution $\tilde{\Lambda}$ and $\varphi _{c\tilde{\Lambda}}$ maximizes the $W$-weighted average power by the Neyman-Pearson lemma, $V_{\tilde{\Lambda}}$ essentially provides an upper bound for the $W$-weighted average power among all tests $\varphi $ that satisfy the uniform size constraint. 

Now, suppose that we obtain some $\tilde{\Lambda}^{\ast }$ on $\Xi $ and the constant $c^{\ast }$ such that 
\begin{equation}
P_{\xi }(\varphi _{c^{\ast }\tilde{\Lambda}^{\ast }}(\mathbf{V}_{\ast})=1)\left. \leq \right. \alpha \text{ for all }\xi \in \Xi \text{,}\label{constraint1}
\end{equation} and
\begin{equation}
\int_{(1-\varepsilon ,2)}P_{\xi }(\varphi _{c^{\ast }\tilde{\Lambda}^{\ast }}(\mathbf{V}_{\ast })=1)dW(\xi )\left. \geq \right. (1-\varepsilon)V_{\tilde{\Lambda}^{\ast }}.  \label{constraint2}
\end{equation}
Then, the test $\varphi _{c^{\ast }\tilde{\Lambda}^{\ast }}$ will have a $W$-weighted average power no less than 100$\varepsilon \%$ lower than any other test of the same level. We set $\varepsilon =0.01$ for our test (\ref{test}). 

The idea of identifying a suitable choice of $\tilde{\Lambda}^{\ast }$ and ${c^\ast}$ is as follows. First, we can discretize $\Xi $ into a grid $\Xi _{a}$ and determine $\tilde{\Lambda}$ accordingly as the point masses. Then we can simulate $N$ random draws of $\mathbf{V}_{\ast }$ from $\xi \in \Xi _{a}$ and estimate $P_{\xi }(\varphi
_{\Lambda }(\mathbf{V}_{\ast })=1)$ by sample fractions. By iteratively increasing or decreasing the point masses as a function of whether the estimated $P_{\xi }(\varphi _{\Lambda }(\mathbf{V}_{\ast })=1)$ is larger or smaller than the nominal level, we can always find a candidate $\tilde{%
\Lambda}^{\ast }$. Note that such $\tilde{\Lambda}^{\ast }$ always exists since we allow $P_{\xi }(\varphi _{\Lambda }(\mathbf{V}_{\ast })=1)<\alpha $ for some $\xi $. We determine $c^{\ast }$ so that (\ref{constraint2}) is satisfied. The continuity of $f_{\mathbf{V}_{\ast }}$ entails that $P_{\xi}(\varphi _{\Lambda }(\mathbf{V}_{\ast })=1)$ as a function of $\xi $ is also continuous. Therefore, (\ref{constraint1}) is guaranteed as we consider  $\left\vert \Xi _{a}\right\vert \rightarrow \infty $ and $N/\left\vert \Xi_{a}\right\vert \rightarrow \infty $, where $\left\vert \Xi _{a}\right\vert $ denotes the cardinality of $\Xi _{a}$. 

In practice, we can determine the point masses by the following concrete steps. 
\\
{\bf Algorithm:}
\begin{enumerate}
\item Simulate $N=$ 10,000 i.i.d. random draws from some proposal density with $\xi $ drawn uniformly from $\Xi _{a}$, which is an equally spaced grid on $[0,0.99]$ with 50 points. 

\item Start with $\tilde{\Lambda}_{(0)}=\{1/50,1/50,\ldots ,1/50\}$ and $c^{\ast }=1$. Calculate the (estimated) coverage probabilities $P_{\xi_{j}}(\varphi _{c^{\ast }\tilde{\Lambda}(0)}(\mathbf{V}_{\ast })=1)$ for every $\xi _{j}\in \Xi _{a}$ using importance sampling. Denote them by $P=(P_{1},...,P_{50})^{\prime }.$ 

\item Update $\Lambda $ by setting $\Lambda _{(s+1)}=\Lambda _{(s)}+\eta_{\Lambda }(P-0.05)$ with some step-length constant $\eta _{\Lambda }>0$, so that the $j$-th point mass in $\Lambda $ is increased/decreased if the coverage probability for $\xi _{j}$ is larger/smaller than the nominal level. 

\item Keep the integration for 500 times. Then, the resulting $\Lambda_{(500)}$ is a valid candidate. 

\item Numerically check if $\varphi _{\Lambda _{(500)}}$ indeed controls the size uniformly by simulating the rejection probabilities over a much finer grid on $\Xi $. If not, go back to step 2 with a finer $\Xi _{a}$. 
\end{enumerate}

%%%%%%%%%%%%%%%%%%%%%%%%%%%%%%%%%%%%%%%%%%%%%%%%%%%%%%%%%%%%%%%%%%%%%

\bibliography{biblio}

\end{document}